\documentclass[12pt,preprint]{aastex}

\bibpunct{(}{)}{;}{a}{}{,}

\slugcomment{nothing}	

\shorttitle{Cyclotron resonance energy variations of 4U~0115+63}
\shortauthors{Nakajima et al.}

\begin{document}

\title{A Further Study of the Luminosity-Dependent Cyclotron Resonance Energies of the Binary X-ray Pulsar 4U~0115+63 with RXTE}

\author{M. Nakajima$^{1,2}$, T. Mihara$^2$, K. Makishima$^{2,3}$ and H. Niko$^3$}
\affil{1) College of Science and Technology, Nihon University,
    1-8-14, Kanda-Surugadai, Chiyoda-ku, Tokyo, JAPAN 101-0062}
\affil{2) Cosmic Radiation, The Institute of Physical and Chemical Research,
    2-1 Hirosawa, Wako, Saitama, JAPAN 351-0198}
\affil{3) Department of Physics, University of Tokyo, 
    7-3-1 Hongo, Bunkyo-ku, Tokyo, JAPAN 113-0033}

\begin{abstract}

The present paper reports on the {\it RXTE} observations of 
the binary X-ray pulsar 4U~0115+63, covering an outburst in 1999
March-April with 44 pointings. 
The 3$-$30 keV PCA spectra and the 15$-$50 keV HEXTE spectra were
analyzed jointly for the cyclotron resonance features.
When the 3$-$50 keV luminosity at an assumed distance of 7 kpc was in 
the range (5$-$13)$\times10^{37}$ erg s$^{-1}$, harmonic double 
cyclotron features were observed in absorption at $\sim$11 and $\sim$22 keV,
as was measured previously during typical outbursts.
As the luminosity decreased below $\sim$5$\times10^{37}$ erg s$^{-1}$, 
the second resonance disappeared, and the fundamental resonance energy 
gradually increased, up to $\sim$16 keV at 0.16$\times10^{37}$ erg s$^{-1}$.
These results reconfirm the report by Mihara et al.\ (2004) using 
{\it Ginga}, who observed a single absorption at $\sim$16 keV in a 
minor ($\sim10^{37}$ erg s$^{-1}$) outburst of this object.
The luminosity-dependent cyclotron resonance energy may be understood as 
a result of a decrease in the accretion column height, 
in response to a decrease in the mass accretion rate.

\end{abstract}


\section{Introduction}

Accreting binary pulsars are considered to have strong surface 
magnetic fields in the range of several times $10^{12}$ G.
One of the methods to accurately measure their fields is to observe 
cyclotron resonance scattering features (CRSFs) 
in their X-ray spectra, because the resonance energy $E_a$
and the magnetic field strength $B$ are related with each other as
$ E_a = 11.6\  B_{12} (1+z_{g}) ^{-1}\ [{\rm keV}] $, 
where $B_{12}$ is the magnetic field strength in units of $10^{12}$ Gauss 
and $z_g$ is the gravitational redshift.
So far, CRSFs have been detected from $\sim$ 15 X-ray pulsars, 
all in absorption, with balloons (e.g., Tr\"umper et al.\ 1978), 
{\it HEAO-1} (Wheaton et al.\ 1979; White et al.\ 1983), 
{\it Ginga} (e.g., Clark et al.\ 1994; Mihara 1995; Makishima et al.\ 1999),
{\it RXTE} (e.g., Coburn et al.\ 2002; Heindl et al.\ 2001),
{\it BeppoSAX} (e.g., Santangelo et al.\ 1999; Orlandini et al.\ 1998), 
and other missions.

The recurrent transient 4U~0115+63, with the 3.6 s pulsations  
(Rose et al.\ 1979), is one of the X-ray pulsars 
whose CRSF has been studied in great detail.
The optical companion is an O9e star, V635 Cassiopeae (Unger et al.\ 1998), 
with the orbital period of 24.3 days. 
The distance to 4U~0115+63 is estimated as 7 kpc 
(Negueruela and Okazaki\ 2001).
Its CRSF was first discovered at $\sim$23 keV in 
the {\it HEAO}-1 A4 spectra by Wheaton et al.\ (1979). 
Using the {\it HEAO}-1 A2 data obtained in the same outburst, 
White et al.\ (1983) suggested that the 23 keV feature is 
in fact the second harmonic resonance, with the fundamental resonance 
at $\sim$11 keV.
The suggested double harmonic structure was reconfirmed with 
{\it Ginga} by Nagase et al.\ (1991). 
Moreover, the third harmonic feature was found with {\it RXTE} 
by Heindl et al.\ (1999), and the fourth harmonic 
with {\it BeppoSAX} by Santangelo et al.\ (1999).
In the {\it BeppoSAX} data, 
the resonance energies were 12.7, 24.2, 35.7 and 49.5 keV, 
or nearly harmonic ratios of 1:1.9:2.8:3.9.
Thus, 4U~0115+63 is one of the most suitable objects 
to study the physics of cyclotron resonance 
in the polar caps of binary X-ray pulsars.

This source has been observed at various X-ray luminosity levels, and
the fundamental resonance energy has been measured repeatedly at $\sim$11 keV.
However, in a minor outburst in April 1991 observed with {\it Ginga},
when the 2$-$60 keV luminosity (1.9$\times10^{37}$ erg s$^{-1}$) was 
$\sim$7 times lower than those typical at the outburst peaks, 
a drastic change in the CRSF was detected: 
instead of the familiar double absorption features at 
$\sim$11 keV and $\sim$22 keV,
it exhibited a single deep and wide absorption at $\sim$16 keV
(Mihara 1995; Mihara et al.\ 1998; Makishima et al.\ 1999; 
Mihara et al.\ 2004, hereafter Paper 1).
Such a large change in $E_a$, presumably depending on the luminosity, 
had been observed neither from 4U~0115+63 itself previously, 
nor from other sources.

In Paper 1, we tentatively concluded that 
the fundamental resonance energy, at $\sim$11 keV in normal outbursts,
increased to $\sim$16 keV, because a lower luminosity 
would make the accretion column shorter, and hence increase 
the magnetic field intensity at the column top.
However, there has remained an alternative possibility that 
the second harmonic resonance, normally at $\sim$22 keV, decreased to 
$\sim$16 keV in the 1991 outburst.
Furthermore, even if the former interpretation is correct, 
the observed change in $E_a$ was considerably larger than is predicted
by an accretion column model by Burnard et al.\ (1991). 
To make the results in Paper 1 less ambiguous, 
we need to more densely sample
the spectra as the luminosity changes over a wide range.

In this paper, we analyzed the {\it RXTE} data of 4U~0115+63 acquired in 
the 1999 March-April outburst, and studied the resonance energy 
as a continuous function of the luminosity 
in the rising and declining phases of the outburst.
We have indeed detected the clear luminosity-dependent changes in the CRSF,
and successfully reconfirmed the inference made in Paper 1.
All of the errors appeared in this paper are 90 \% confidence levels.

\section{Observations}

\begin{figure}
\plotone{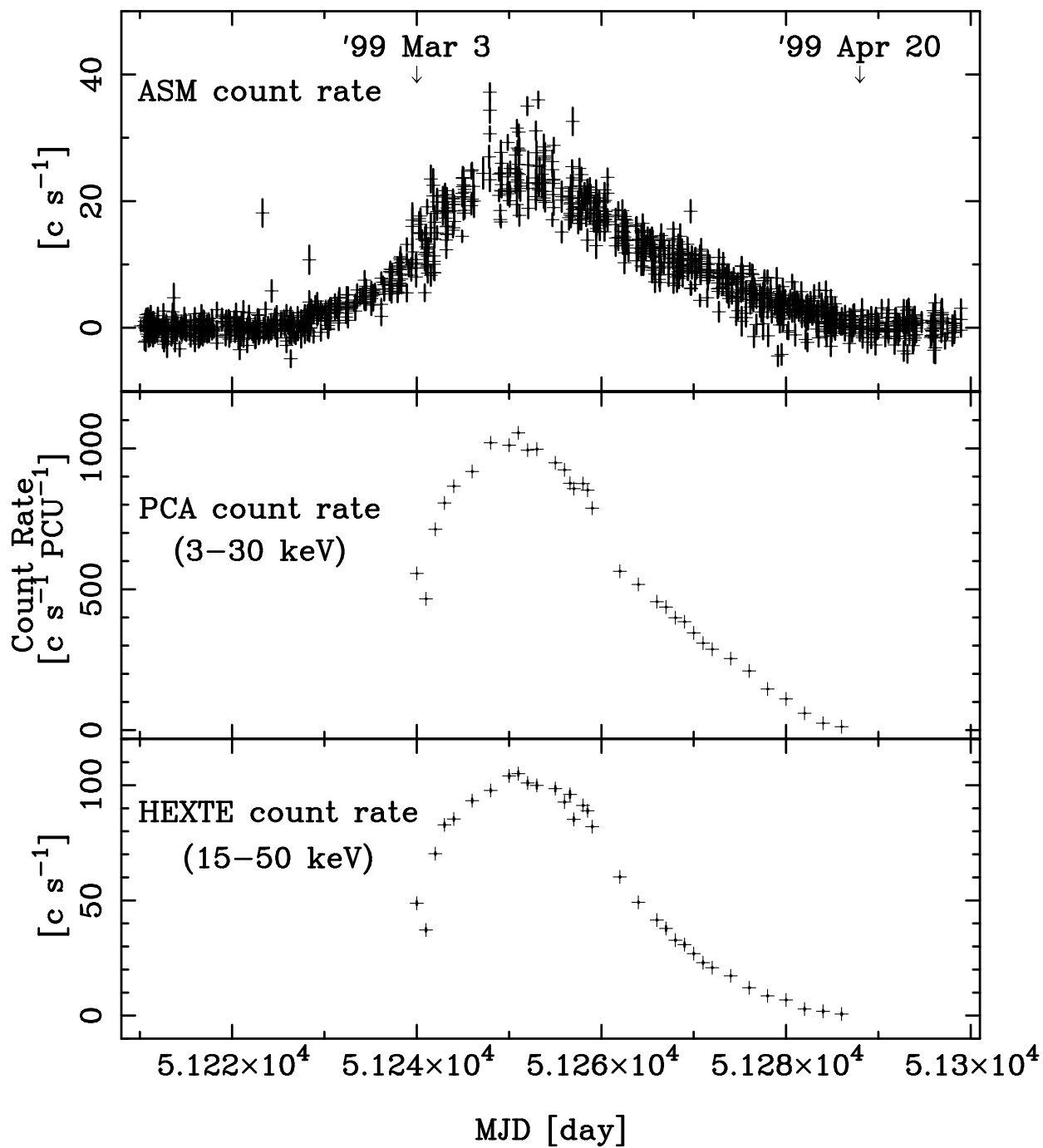}
\caption{(top) The 2$-$10 keV {\it RXTE} ASM light curve of 4U~0115+63 in the 1999 March-April outburst. (middle and bottom) The 3$-$30 keV PCA and the 15$-$50 keV HEXTE light curves obtained from the 33 pointing observations listed on Table \ref{tab1}. }
\label{fig1}
\end{figure}

The outburst of 4U~0115+63 to be utilized in the present paper 
was detected in 1999 March with 
the All Sky Monitor (ASM) on board {\it RXTE}.
Figure \ref{fig1} shows the light curve of the whole outburst.
The peak intensity, $\sim$30 c s$^{-1}$, corresponds to approximately 
400 mCrab, which is typical of regular outbursts of this source.
The {\it BeppoSAX} observation of this outburst led to 
the discovery of the four harmonic CRSFs by Santangelo et al.\ (1999).

During this outburst, 44 short ($\sim1$ ksec) pointing 
observations were made with 
the Proportional Counter Array (PCA; Jahoda et al.\ 1996) and 
the High-Energy X-ray Timing Experiment (HEXTE; Rothschild et al.\ 1998)
on board {\it RXTE}, starting on 1999 March 3 and ending on April 20.
Heindl et al.\ (1999) utilized these datasets, and
successfully detected up to three CRSFs. 
The PCA consists of five xenon proportional counter units (PCUs) with an 
energy range of 2$-$60 keV and a total effective area of $\sim$7000 cm$^2$.
In order to extend the detector life time, some of the PCUs were not 
operated in the present observations.
In this paper, we utilize all PCUs operated at each observation,
as summarized in Table \ref{tab1}.
The HEXTE consists of two arrays (cluster A and B) of four NaI/CsI 
scintillation counters, with an energy range of 15$-$250 keV and a total 
effective area of $\sim$1600 cm$^2$.
However, one pulse height analyzer of cluster B failed in the early mission 
phase, so its effective area is approximately 3/4 of the nominal value.

From the 44 pointing observations of the 1999 outburst 
in the {\it RXTE} archive, we selected 34 data with 
low electron contamination; these are listed in Table \ref{tab1}.
In this selection, we set the electron-rate threshold at 
0.12 counts s$^{-1}$ (which is a typical value when the signal X-ray flux 
is high), in order not to discard good data by 
mistaking signal X-rays for electrons.
In order to complement the rather high threshold,
we visually inspected the electron-rate light curves,
and confirmed that none of the 34 datasets are contaminated 
by sporadic electron events. 
We then selected the Standard-2 mode data which have 16 s
time resolution and 129 energy channels, and
used only the top layer of the PCA which has a low background.
We employed FTOOLS v5.2 for analysis, and 
used the calibration data file epoch 3 and 4 
to make the energy response matrix.
The background subtraction utilized so-called 
bright background model for all data.
We have selected those data acquired when 
the offset angle of the source to the field of view center was 
$<0^{\circ}.02$, the earth elevation angle was $>10^{\circ}$, 
and the spacecraft was not within 30 minutes of an entrance 
from the South Atlantic Anomaly.
The net exposure of each datasets, which was obtained after
these screenings, is also given in Table \ref{tab1}.

Since different observations used different sets of PCUs (Table \ref{tab1}),
it is important to evaluate systematic errors associated with 
the PCA responses (particularly among different PCUs).
Accordingly, we followed Wilms et al.\ (1999) and Coburn et al.\ (2001),
and analyzed the Crab nebula data obtained 
on 1997 December 12 (for epoch 3) for $2.8\sim5.4$ ksec, 
and 1999 December 18 (for epoch 4) for $1.9\sim8$ ksec,
with the exposure depending on PCUs.
By fitting these data with a single power-law of photon index 2.1,
which is known to be a good approximation of the Crab spectrum 
in the PCA range,
we examined how the data-to-model ratio at each energy bin 
scatters among the PCUs.
Form this, we have set the systematic error of the PCU to be 1 \%.
The subsequent results do not change if we instead use 0.5 \%.

\section{Analysis and Results}

\subsection{Ratios to the Crab spectrum}

\begin{figure}
\plotone{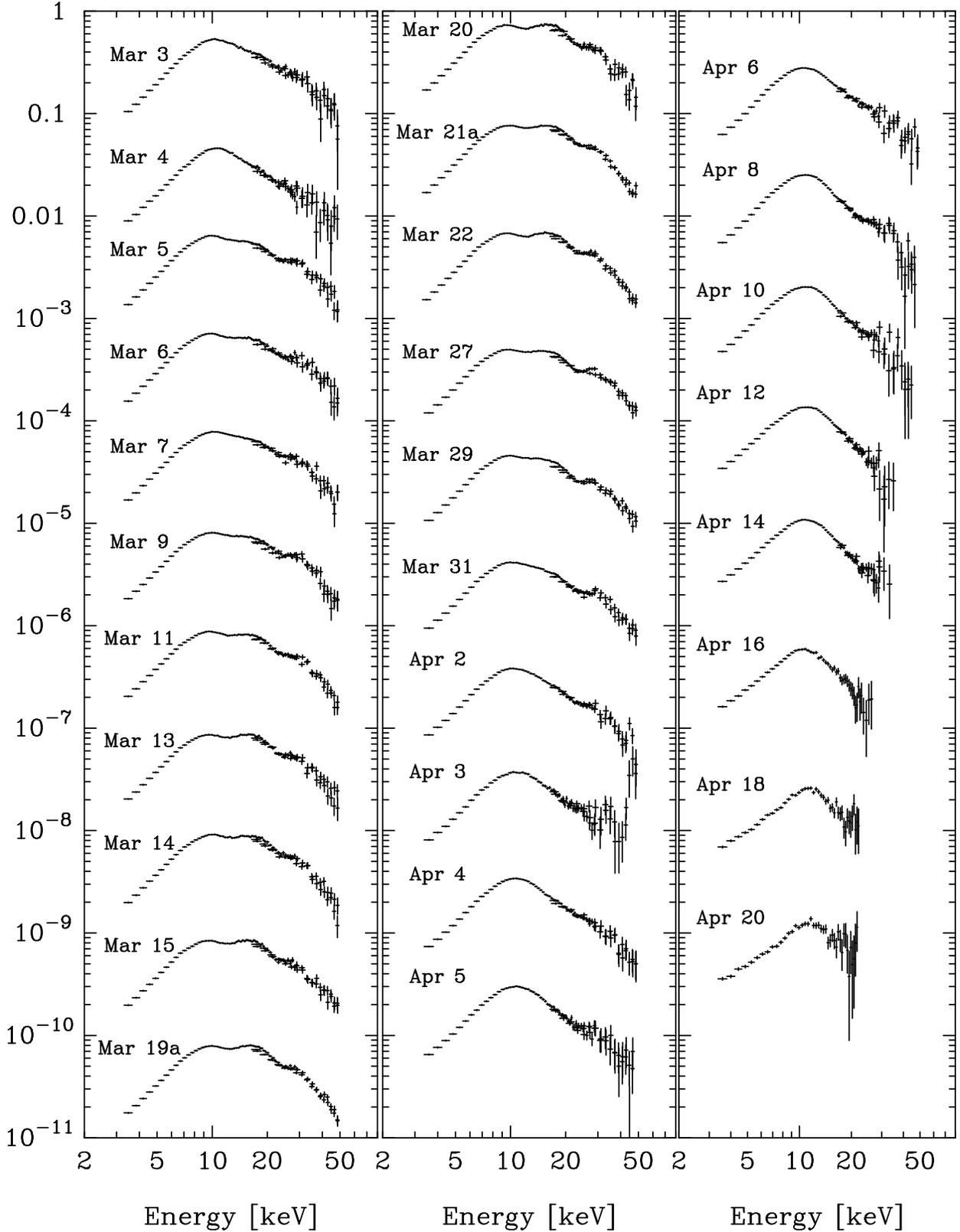}
\caption{Background-subtracted PCA and HEXTE spectra of 4U~0115+63 in the 1999 outburst, shown normalized to those of the Crab Nebula measured by respective instruments. Above plot exclude the same observation date(March 19b, 21b), low exposure data (March 16, 18 and April 1).}
\label{fig2}
\end{figure}

For each dataset listed in Table \ref{tab1}, we accumulated 
3$-$30 keV PCU data into a single spectrum, and
15$-$50 keV HEXTE data into another.
The background was subtracted in the way described in section 2.
In order to grasp the spectral shape in a model-independent manner 
without being hampered by the instrumental response, 
we surveyed these spectra normalized to that of the Crab spectrum.
The reference Crab data are the same as utilized 
to estimate the systematic error.
The results are shown in Figure \ref{fig2}, where
errors which propagated from the Crab spectrum are negligible.

In Figure \ref{fig2}, the first data set acquired on March 3 
gives a relatively featureless spectral ratio.
When the flux increased and exceeded 710 c s$^{-1}$ (PCU)$^{-1}$
on March 5, the Crab ratio started to reveal 
two dip features at $\sim$12 keV and $\sim$25 keV.
These can be interpreted as the fundamental and second 
harmonic, as have been observed repeatedly.
The double absorption feature persisted throughout the flare peak 
until March 29.
However, when the flux decreased to $\leq$400 c s$^{-1}$ (PCU)$^{-1}$ 
on April 2, it turned into an apparently single absorption feature.
Thus, the present data appears to reconfirm 
the {\it Ginga} results reported in Paper 1,
that the CRSF appears in double absorption when the X-ray intensity is high,
while it changes into a single absorption as the source becomes faint. 
In the {\it RXTE} data, the threshold is suggested to lie
between 400 and 710 c s$^{-1}$ (PCU)$^{-1}$.
Below, we attempt to quantitatively confirm these inspections.

\subsection{Analysis of representative spectra}

It is known that typical continuum spectra of accreting binary pulsars 
can be approximated with a power-law times exponential cutoff model.
In this paper, we employ its updated version called 
NPEX (Negative and Positive power-laws with EXponential) model
(Mihara 1995; Makishima et al.\ 1999),
and fit the background-subtracted spectra 
in comparison with the results reported in Paper 1.
The NPEX model is written as
\begin{equation}
NPEX(E) = ( A_1 E^{-\alpha_1} + A_2 E^{+\alpha_2} ) ~ \exp \left( -\frac{E}{kT} \right),
\label{eq1}
\end{equation}
where $E$ is the X-ray energy in units of keV, $A_1$ and $\alpha_1$ are 
the normalization and photon index of the negative power-law, respectively, 
$A_2$ and $\alpha_2$ are those of the positive power-law, and 
$kT$ represents the cutoff-energy in units of keV.
In this paper, we tentatively fix $\alpha_2$ at 2.0,
so that the positive power-law describes a Wien peak
(Rybicki \& Lightman 1979).

\begin{figure}
\plotone{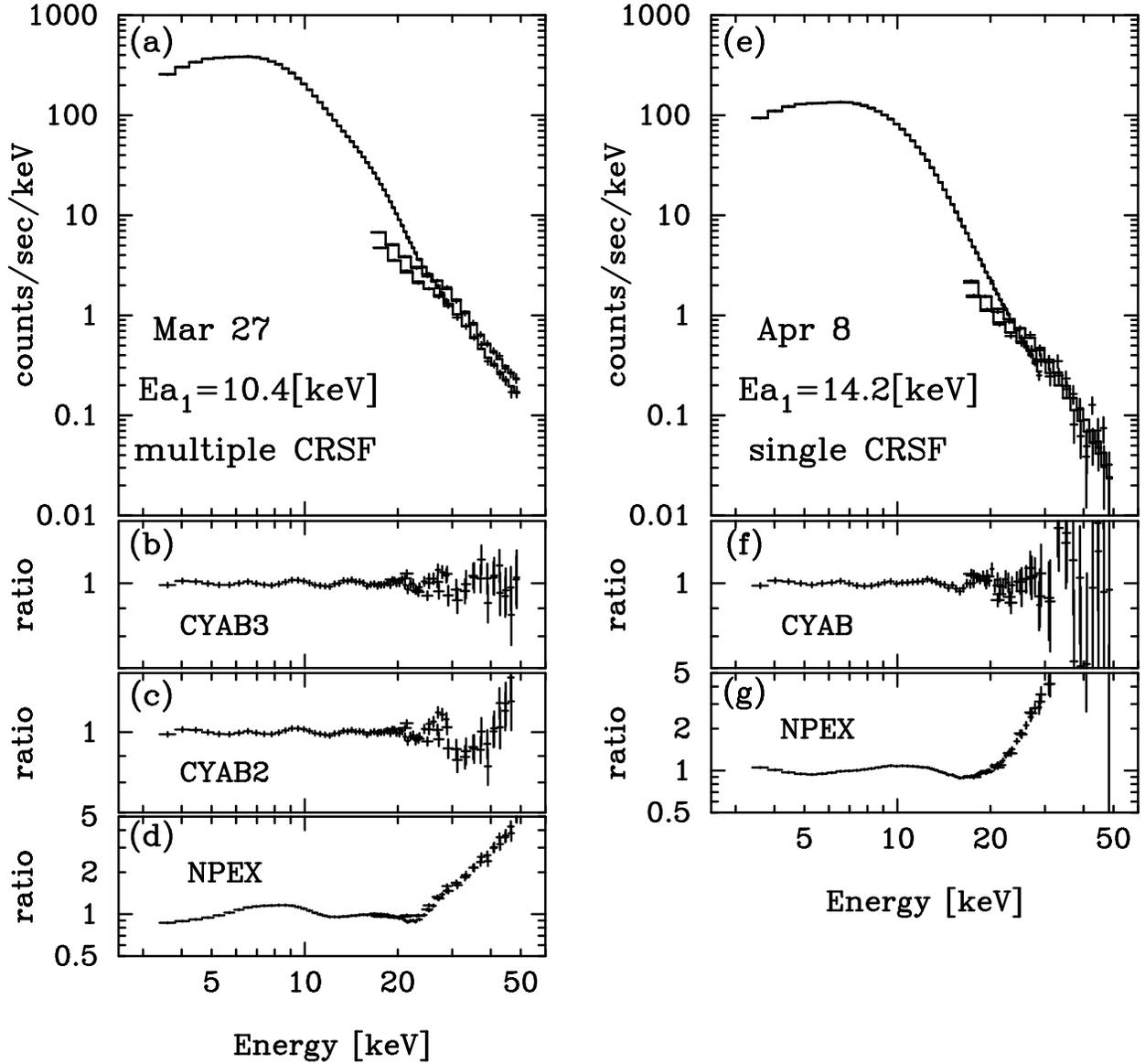}
\caption{Pulse-phase-averaged spectra of 4U~0115+63 on March 27 (left) and April 8 (right).  The top panels ((a) and (e)) are the background-subtracted spectra of the PCA and HEXTE, shown in comparison with the best-fit NPEX$\times$CYAB model (histograms). Two resonances are incorporated to fit the March 27 data, while a single resonance for the April 8 data. The ratios from the NPEX model fits are shown in the bottom panels ((d) and (g)), whereas those from the NPEX model with triple (b), double (c) and single (f) CRSF are shown in the middle panels.}
\label{fig3}
\end{figure}

Figure \ref{fig3} shows the pulse-phase-averaged PCA$+$HEXTE spectra 
obtained on March 27 and April 8, 
when the Crab ratio suggests the double and single CRSF, respectively 
(see Figure \ref{fig2}).
We first attempted to fit the spectra with the NPEX model.
In order to take into account possible over- or under-subtraction 
of background, we allowed the background normalization to vary
(as described in the {\it RXTE cook book}), 
so as to minimize the fit chi-squared.
We found the optimum normalization factor to be close to 
the nominal value of 1.0 within a few percent.
Even allowing this correction, the model left 
significant structures in data-to-model ratios as shown 
in Figure \ref{fig3}d and g. 
As a result, the fits remained unacceptable with $\chi_{\nu}^{2} \sim 62$ and
$\chi_{\nu}^{2} \sim 30$ for the March 27 and April 8 spectra, respectively.
Indeed, the March 27 spectrum exhibits two negative deviations at 
$\sim$12 and $\sim$22 keV from the NPEX fit, 
while that of April 8 shows only one negative feature around 16 keV.
These results reconfirm the inference from the Crab ratios.

In order to better reproduce the spectra, 
we next introduced a cyclotron absorption (CYAB) factor
which is given as
\begin{equation}
CYAB(E) = \exp{ \biggl\{ - \frac{\tau~(WE/E_a)^2}{(E-E_a)^2+W^2} \biggl\} },
\label{eq2}
\end{equation}
where $E_a$ is the resonance energy, $W$ is the width of 
the absorption structure, and $\tau$ is the depth of the resonance 
(Mihara et al.\ 1990).
We fitted the spectrum on March 27 with the NPEX continuum multiplied 
by two CYAB factors (hereafter NPEX$\times$CYAB2 model),
in which all the NPEX and CYAB parameters are left free except 
that the second CRSF energy is fixed at twice the fundamental energy.
For the spectrum on April 8, we applied the NPEX multiplied by
a single CYAB factor (hereafter NPEX$\times$CYAB model).
The PCA$+$HEXTE data were fitted simultaneously with the same parameters,
but using another free parameter to adjust relative normalizations of 
the two instruments.
The background normalization was again allowed to vary, 
by up to $\pm1.5$ \%.
We did not include low energy absorption nor Fe-K line, 
since our spectra exhibit neither low-energy turn-offs nor
an excess around 6.4 keV.
The lack of significant Fe-K line is often noticed among  transient
X-ray pulsars with Be-type primary stars (e.g., Nagase 1989);
in such a system, the matter to be accreted by the pulsar
presumably forms a disk-shaped envelope around the Be star,
and hence its solid angle as seen from the pulsar is much smaller
than the case where the matter is captured from a more isotropic stellar wind.

As shown in Figure \ref{fig3}e and f, 
the NPEX$\times$CYAB model has successfully reproduced 
the PCA$+$HEXTE data on April 8, yielding 
reduced chi-squared of $\sim$1.2.
The background correction factor turned out to be 0.3 \%.
In contrast, the March 27 spectrum was not reproduced successfully 
($\chi_{\nu}^{2} \sim 2.0$) even by the NPEX$\times$CYAB2 model,
because of negative residuals around 35 keV (Figure \ref{fig3}c).
This feature was removed by multiplying a third CYAB factor, 
of which the centroid energy is $32.8_{-8.3}^{+3.2}$ keV.
Since this energy is close to $3\times E_{a1}$,
the feature can be identified with the third harmonic resonance 
detected from the same outburst
(Heindl et al.\ 1999; Santangelo et al.\ 1999).
We hence fitted the March 27 spectrum with the NPEX continuum
multiplied by three CYAB factors, of which 
the resonance energies were constrained to have harmonic ratios
($1:2:3$).
This model, hereafter called NPEX$\times$CYAB3 model, 
successfully reproduced the March 27 spectrum 
with $\chi_{\nu}^{2} \sim 1.2$.
The derived fundamental resonance energy on March 27 is $10.4 \pm 0.1$ keV 
at an X-ray luminosity of 7.6$\times 10^{37}$ erg s$^{-1}$, while that of 
April 8 is $14.2 \pm 0.4$ keV at 3.0$\times 10^{37}$ erg s$^{-1}$.
The resonance energy has thus increased by a factor of 1.4 
as the luminosity decreased by a factor of 2.5, 
although there still remains a possibility that 
the single CRSF at 14.2 keV on April 8 is in reality the second harmonic.

While we tentatively fixed $\alpha_2$ at 2.0,
Santangelo et al. (1999) obtained $\alpha_2 = 0.41 \pm 0.05$
by analyzing the {\it BeppoSAX} data of the same outburst.
This is probably due to difference in the energy band used;
we fitted the spectra in the $3-50$ keV range,
while Santangelo et al.\ (1999) in $9-100$ keV. 
To examine this difference,
and to examine whether the constraint of $\alpha_2 = 2.0$ is justifiable,
we fitted the NPEX $\times$ CYAB3 model, with $\alpha_2$ left free,
to the PCA$+$HEXTE data on March 19a,
which is one of those with the highest statistics.
Then, this parameter has been constrained as $\alpha_2 = 1.60 \pm 0.99$,
which includes the fiducial value of 2.0.
More specifically,
fixing $\alpha_2$ at 2.0, at our best fit value of 1.60,
and at the {\it BeppoSAX} value of 0.41,
gave a reduced chi-squared of
1.29 ($\nu=78$) , 1.33 ($\nu=77$) , and 1.67  ($\nu=78$), respectively.
Based on these results,
we hereafter retain our initial assumption of $\alpha_2 = 2.0$.

Apart from the continuum modeling, we must examine
whether the CRSF parameters we have derived are consistent 
with the previous analyses of the same outburst
(Heindl et al.\ 1999; Santangelo et al.\ 1999).
We therefore analyzed the PCA$+$HEXTE spectra
on March 11.36$-$11.40 (Obs. 7; Table \ref{tab1}),
in comparison with those reported by Heindl et al.\ (1999)
for a particular pulse phase (0.70$-$0.76 in terms of their phase convention)
on March 11.87$-$12.32.
We found some difference in the resonance energy;
$10.6_{-0.2}^{+0.1}$ keV from our fit, 
while $12.40_{-0.35}^{+0.65}$ keV by Heindl et al.\ (1999).
The difference is likely to arise from the different modeling
of the CRSF employed in these studies,
because Heindl et al.\ (1999) used a Gaussian absorption model
whereas we are using the CYAB factor.
Accordingly,
we re-fitted the same spectra 
with an NPEX model multiplied by two Gaussian absorption models.
The fundamental cyclotron resonance was
then obtained at $E_{a1} = 12.3\pm0.1$ keV,
which is consistent with  Heindl et al.\ (1999).
The higher accuracy of our result is probably
because we analyzed the pulse phase-averaged spectra.
Other resonance parameters turned out to be consistent
between the two works.
We thus conclude that our results are consistent
with those of Heindl et al.\ (1999) using the longer exposure,
and have comparable accuracy.

\subsection{Analysis of date-sorted spectra}
\label{sec:3.3}

\begin{figure}
\plotone{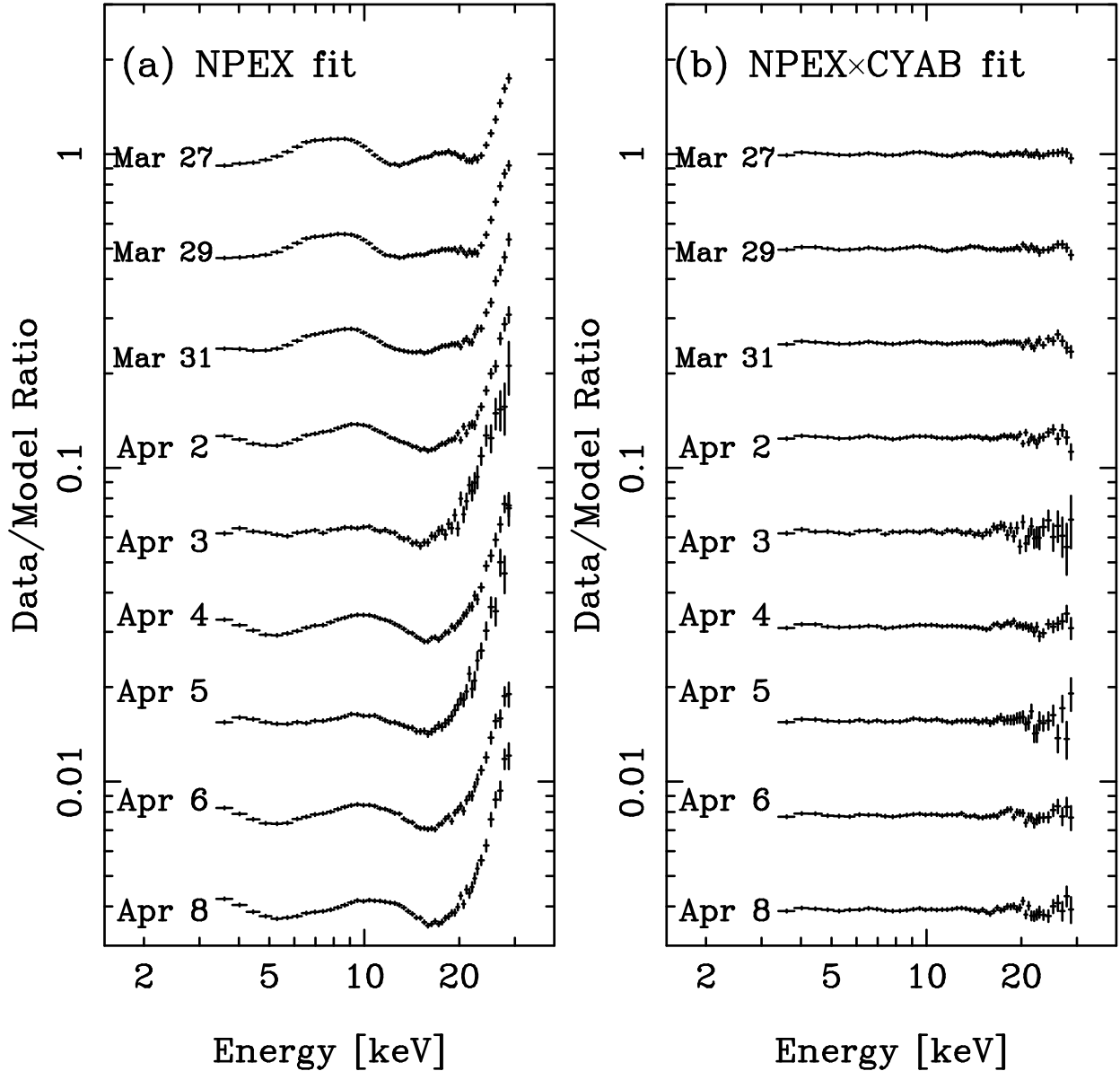}
\caption{(a) Date-sorted PCA spectra of 4U~0115+63 from March 27 to April 8, normalized to the respective best-fit NPEX models. For the presentation, the results are shifted vertically by a factor of 0.5 for each observation.  (b) The same as panel (a), but the fitting model is NPEX$\times$CYAB2 (March 27 through April 2) or NPEX$\times$CYAB (April 4 and later).}
\label{fig4}
\end{figure}

For the moment, 
we mainly concentrate on the study of the 3$-$30 keV PCA spectra, 
because this energy range is most relevant to the CRSFs.
In order to more continuously resolve the drastic spectral change 
between March 27 to April 8,
we applied the NPEX model (without incorporating CYAB factors) to
the daily-averaged PCU0$+$PCU2$+$PCU3 spectra over this period.
Figure \ref{fig4}a shows ratios of the data to the best-fit 
NPEX model, which allow us to grasp the spectral changes objectively.
The PCA background normalizations were corrected in the way 
as described in section 3.2.
Since the ratio on March 27 and 29 shows almost the same shape 
as Figure \ref{fig3}d, 
we may consider that the two CRSF persisted at least until March 29.
On March 31, the second CRSF became less clear, and
on April 2 onward, the ratio reveals only a single absorption.
In addition, the fundamental resonance initially observed at $\sim11$ keV
moved to higher energies, up to $\sim15$ keV, 
while there is no opposite trend such as 
the two features drifting toward lower energies.

In order to quantify the results of these visual inspections,
we have fitted the same set of spectra 
with the NPEX$\times$CYAB2 (or NPEX$\times$CYAB) model.
The obtained best-fit parameters are summarized in Table \ref{tab2},
and the ratio of the data to the model are plotted in 
Figure \ref{fig4}b.
The spectra from March 27 through April 2 all required 
the two CYAB factors,
because the double CYAB fit gave a significantly better reduced 
chi-square (0.6$\sim$1.2) than the single CYAB case ($\geq$2).
On April 4, the spectrum is roughly reproduced by the single CYAB model,
although there remains a hump at $\sim$23 keV which does not disappear by
including the second CYAB factor.
The successful single CYAB fit continued to the end, on April 8.
These results quantitatively confirm our inference made above,
that the fundamental CRSF energy increased 
from 10 to 15 keV over this $\sim$10 day period as assumed in Paper 1, 
although details of the change are not yet resolved clearly.

\subsection{Analysis of intensity-sorted spectra}
\label{sec:3.4}

\begin{figure}
\plotone{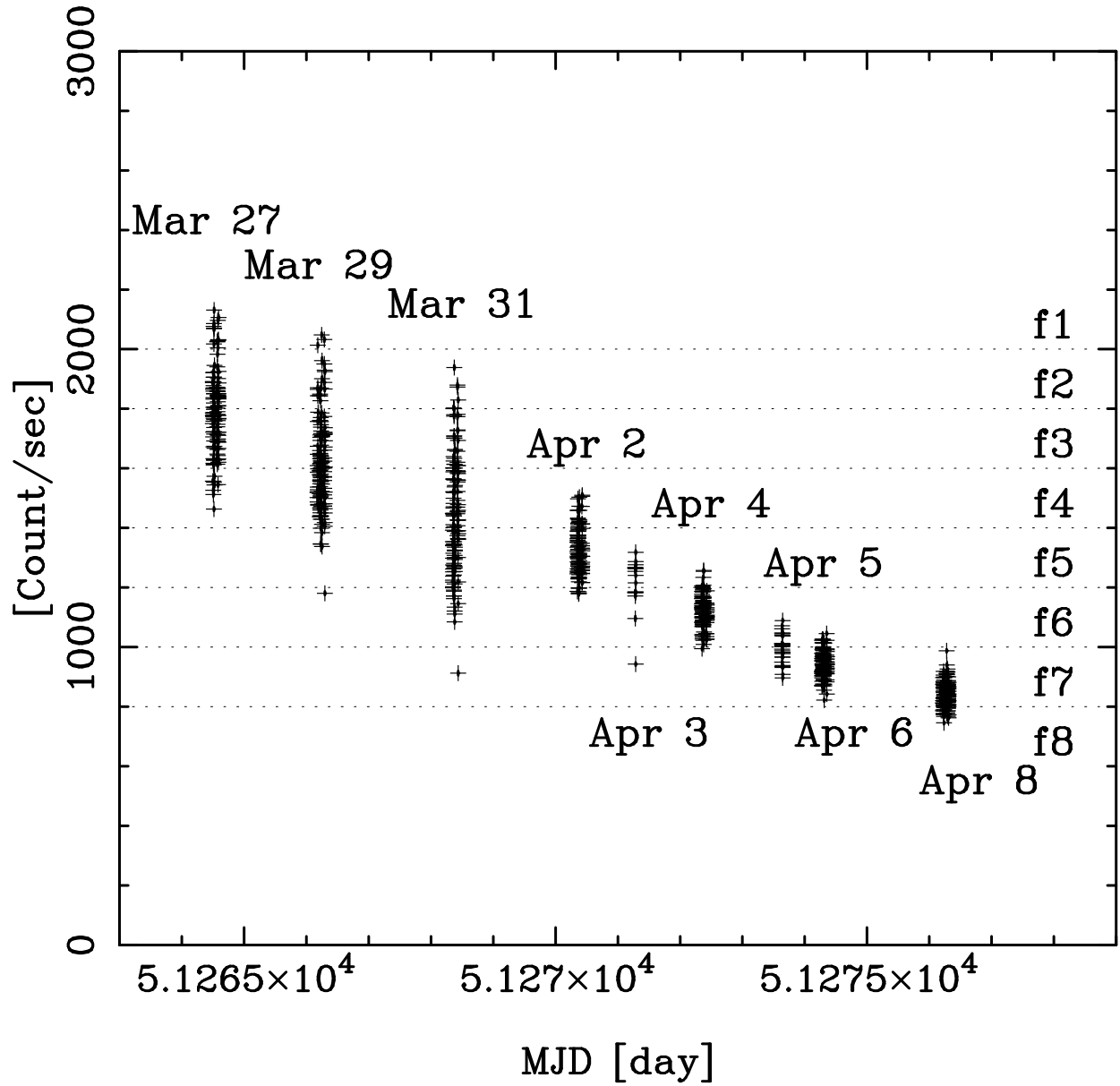}
\caption{The PCU $0+2+3$ light curve of 4U~0115+63 observed from March 27 to April 8, plotted with 16 sec binnings.  The horizontal dashed lines indicate boundaries of the intensity sorting.}
\label{fig5}
\end{figure}

Although the 3$-$30 keV spectrum apparently depends on the
luminosity, the source exhibits significant intra-day variations, 
as shown in Figure \ref{fig5}.
As a result, each spectrum in Figure \ref{fig4},
in reality, is an average over a relatively wide luminosity range.
In order to see the luminosity dependence more clearly, 
we sorted the PCA data from March 27 to April 8 into 8 intensity intervals,
in reference to Figure \ref{fig5}.
We again used data from PCU 0, 2 and 3, which worked throughout this period.
Then, we have repeated the same analysis as performed 
in section \ref{sec:3.3}.

\begin{figure}
\plotone{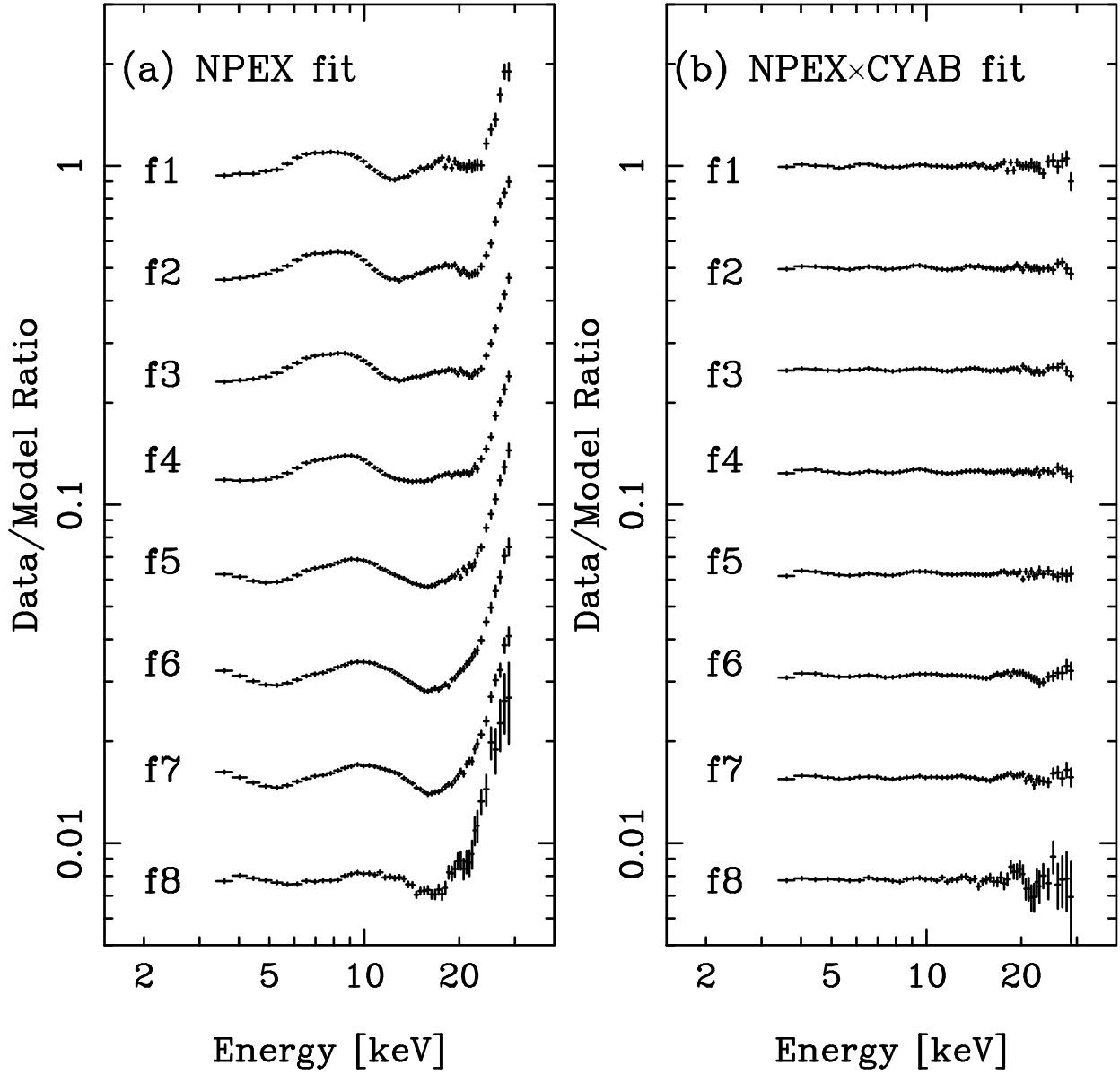}
\caption{(a) The same as Figure \ref{fig4}a, but for the intensity-sorted spectra defined in Figure \ref{fig5}.   (b) The intensity-sorted PCA spectra, each normalized to the best-fit NPEX$\times$CYAB2 (f1-f5) or NPEX$\times$CYAB (f6-f8) model. Data are presented in the same manner as Figure \ref{fig4}.}
\label{fig6}
\end{figure}

Figure \ref{fig6}a shows the ratios of these intensity-sorted
spectra to their best-fit NPEX models,
to be compared with Figure \ref{fig4}a.
The two CRSFs, at $\sim$11 keV and $\sim$22 keV, 
are thus observed clearly in the higher four intensity levels, f1 through f4.
As the intensity decreased, 
the second CRSF gradually became shallower.
Finally, the second CRSF disappeared at level f6, and
the fundamental CRSF started to move 
from $\sim$11 to $\sim$16 keV over levels f5 through f8.
From these results, we confirm that the $\sim$16 keV single structure 
results from an upward shift of the $\sim$10 keV fundamental CRSF.
The spectral changes revealed here are consistent with, but clearer than,
those seen between April 2 and April 4 in the date-sorted spectra.

\begin{figure}
\plotone{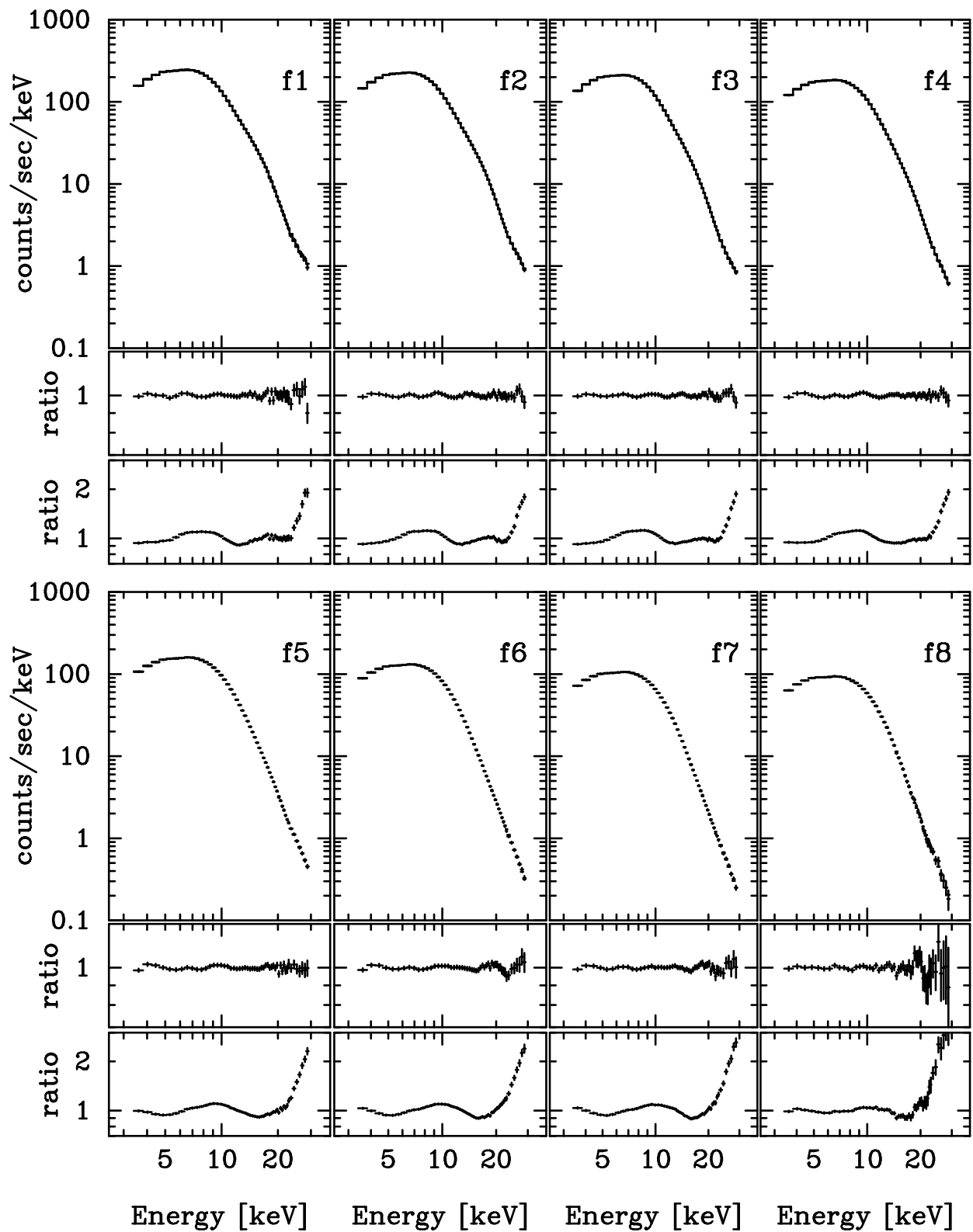}
\caption{The eight intensity-sorted spectra defined in Figure \ref{fig5}, fitted with the NPEX$\times$CYAB2 (f1-f5) or the NPEX$\times$CYAB (f6-f8) model. The middle panels show the data to model ratios, while the bottom panels are the same ratios but without incorporating the CYAB factor(s). }
\label{fig7}
\end{figure}

Using the NPEX$\times$CYAB2 (or NPEX$\times$CYAB) model,
we quantified the CRSF parameters as a function of the intensity.
The fitting results are summarized in Table \ref{tab3}, and
the ratios of the data to the model are shown 
in Figure \ref{fig6}b.
The full fitting results are displayed in Figure \ref{fig7}.
The two CYAB factors have been required by the f1 to f4 spectra.
The second CRSF is not clearly visible in the f5 ratio 
(Figure \ref{fig6}a and Figure \ref{fig7}), 
but a large $\chi^2_{\nu}$ ($\sim$2.0) was obtained by 
the single CYAB fitting.
We therefore applied the NPEX$\times$CYAB2 model to the f5 spectrum,
and obtained a fully acceptable fit (Table \ref{tab3}).
This means that the second harmonic is still present in the f5 spectrum, 
with a reduced depth ($\tau_2 \simeq 0.2$).
From f1 through f5, both $\tau_2$ and $W_2$ decreased; 
that is, the second harmonic feature became narrower and shallower.
Although this behavior is absent in the day-sorted results 
(Table \ref{tab2}), the difference can be attributed to the fact that 
each day-average spectrum is a mixture of different spectra 
corresponding to different intensities.
The f6 through f8 spectra have been fitted successfully by 
the NPEX$\times$CYAB model.

These results unambiguously show that the center energy of 
the fundamental CRSF moved from 10 to 15 keV as the luminosity decreased.
Thus, we can conclude that the 16 keV spectral structure of 4U~0115+63 
observed by {\it Ginga} (1991) and this work is the fundamental CRSF,
rather than the second harmonic resonance which moved toward lower energies.
The threshold between the single and double CRSF 
structures is found at a $3-30$ keV luminosity of
$\sim$4.2$\times10^{37}$ erg s$^{-1}$.

\subsection{Analysis of all PCA and HEXTE spectra}
\label{sec:3.5}

Now that we have understood the basic behavior of the CRSFs, 
we now proceed to the analysis of the entire PCA and HEXTE data 
prepared in section 3.1.
Since the Crab ratios on March 3 and 4 do not show clear CRSFs
(Figure \ref{fig2}),
we first attempted an NPEX fitting,
but the model was unacceptable ($\chi^2_{\nu} \geq 20$), 
because of the negative deviation at $\sim15$ keV.
Then we applied the NPEX$\times$CYAB model to these data, 
but the fit was still unacceptable ($\chi^2_{\nu} \geq 1.7$).
Finally, we fitted these data with the NPEX$\times$CYAB2 model, and 
obtained reasonable fit with $\chi^2_{\nu} \leq 1.3$.
These results imply that the double CRSFs were already present 
in these early data sets, with the fundamental resonance at $\sim12$ keV
as given in Table \ref{tab4}.
Thus, the data from the beginning through April 2 
were generally well described with the NPEX$\times$CYAB2 model.
By jointly analyzing the PCA and HEXTE data, 
the second CRSF parameters have been significantly better constrained than 
using the PCA data alone.
The derived parameters are listed in Table \ref{tab4}.

In some fits, however, 
we were still left with rather large values of $\chi^2_{\nu}$.
For example, the NPEX$\times$CYAB2 fits are rather poor 
($\chi^2_{\nu} \geq 1.5$) on March 15, 19a, 21a, 22, 27, 29 and 31
(Table \ref{tab4}), 
often with negative residuals at $\sim$35 keV. 
Like in Figure \ref{fig3}b, 
these residuals can be mostly removed by multiplying 
the model with a third CYAB factor.
Typical parameters of the third CRSF, e.g., observed on March 19a,
are $E_{a3}=35.2\pm2.2$ keV, $W_3=9.3\pm2.5$ keV and $\tau_3=0.6\pm0.1$.
We can hence identify it with the third harmonic, after section 3.2.
The third harmonic persisted in the spectra for about half a month 
from March 15, with relatively constant parameters, till March 31
when it became unconstrained presumably due to 
insufficient data statistics in higher energies.
Meantime, the second CRSF stayed at around 22 keV.

The data from April 3 to the end did not show the higher harmonics, 
with the NPEX$\times$CYAB model giving acceptable fits ($\leq 1.4$).
In order to estimate the upper limits on the second harmonic feature,
we attempted to fit the data obtained from April 3 to April 8
with the NPEX$\times$CYAB2 model. 
In addition to the constraint of $E_{a2} = 2 E_{a1}$ employed so far,
the second CRSF width $W_2$ was tied to $W_1$,
because $W_2$ is close to $W_1$ (except for April 1 and 2)
when the spectrum exhibits the double features.
The obtained upper limits on $\tau_2$ are given in Table \ref{tab4}.

From these studies using the PCA and HEXTE data, 
we reconfirmed that the cyclotron resonance energies 
increase as the X-ray luminosity decreases.
These results are consistent with those derived in section 3.4 from 
the intensity-sorted study of the $3-30$ keV PCA spectra.
Figure \ref{fig8} summarizes the fundamental resonance energies,
derived through both the date-sorted and intensity-sorted analyses, 
as a function of the calculated $3-50$ keV luminosity at 7 kpc.

\section{Discussion}
\label{sec:4}

We have analyzed the 34 PCA$+$HEXTE datasets,
covering the whole 1999 March-April outburst of 4U~0115+63
in which the $3-50$ keV source luminosity changed over
$L_{\rm x} = (0.17-14) \times 10^{37}$ erg s$^{-1}$.
When $L_{\rm x} > 7 \times 10^{37}$ erg s$^{-1}$,
we observed the familiar double CRSFs with $E_{\rm a1} \simeq 11$ keV.
As $L_{\rm x}$ decreased across a rather
narrow range of $( 5 \pm 2) \times 10^{37}$ erg s$^{-1}$,
the second harmonic resonance disappeared,
and the fundamental resonance energy increased
from $E_{\rm a1}\sim 11$ keV to $\sim 16$ keV.
These results reconfirm Paper 1,
and unambiguously reveal
that the resonance energy {\it increases}
as the source gets less luminous.

\subsection{The resonance energy variations}

Except for some hysteresis effects
between the rising and decay phases of the outburst,
$E_{\rm a1}$ changes roughly as a single-valued function of $L_{\rm x}$.
In particular, the intensity-sorted 3$-$30 keV PCA spectra
and the date-sorted PCA$+$HEXTE data imply consistent results.
Furthermore, the {\it Ginga} results reported in Paper 1
fall on the same $E_{\rm a1}$ vs. $L_{\rm x}$ relation.
Therefore, the phenomenon is inferred to have a good reproducibility
as a function of the luminosity.

The CRSF is known to depend also on the pulse phase, 
presumably because different scattering regions characterized 
by slightly different magnetic field strengths are observed 
as the neutron star rotates.
When studying luminosity-related changes in the resonance energy,
such pulse-phase dependent effects must be considered.
In the case of 4U~0115+63, however,
the phase dependent variation of $E_{\rm a1}$ in the 1999 outburst
was relative small, $\sim10$\% (Santangelo et al.\ 1999).
Furthermore, the phase-resolved analysis performed in Paper 1, 
on the two different luminosity levels, did not affect the main results 
from the phase-averaged spectroscopy.
We therefore concentrate, in the present paper, 
on the phase-averaged analysis.

As mentioned in Paper 1,
luminosity-dependent changes of the accretion column height
provide the most likely explanation to our results,
because the column is expected to become taller
as the source luminosity increases (e.g., Burnard et al. 1991).
Assuming that the CRSF is formed at a height $h_{\rm r}$
above the neutron star surface in the accretion column,
and the magnetic field strength there follows the dipole law,
we then expect
\begin{equation}
E_{\rm a1} \propto (R_{\rm NS} + h_{\rm r})^{-3} (1+z_{\rm g})^{-1}
\label{eq3}
\end{equation}
with $R_{\rm NS}$ the radius of the neutron star.
The factor $(1+z_{\rm g})$ describes the gravitational redshift,
but below we assume $z_{\rm g} =0$ for simplicity.
Then, the relative resonance height, $h_{\rm r}/R_{\rm NS}$,
can be related to the resonance energy as
\begin{equation}
  \frac{h_{\rm r}}{R_{\rm NS}} \approx \
  \left( \frac{E_{\rm a}}{E_{\rm 0}} \right)^{-1/3} - 1 ~,
\label{eq4}
\end{equation}
where $E_{\rm 0}$ is the resonance energy
to be observed on the neutron star surface.

Substituting the observed value of
$E_{\rm a1}$ into Equation (\ref{eq4}),
we have calculated $h_{\rm r}/R_{\rm NS}$
as a function of the X-ray luminosity.
The results are shown in Figure \ref{fig9}.
Since there is no {\it a priori} knowing of $E_0$,
we employed two different values of $E_0$;
18 keV and 20 keV,
the former close to the observed maximum value of $E_{\rm a1}$.
Except for some differences between the two assumptions on $E_0$,
the height $h_{\rm r}$ thus increases
in a rough proportion to the X-ray luminosity,
up to $\sim 7 \times 10^{37}$ erg s$^{-1}$
where $h_{\rm r}$ appears to saturate at $\sim 0.2 R_{\rm NS}$.
In addition,  $h_{\rm r}/R_{\rm NS}$ may approach a finite value of 
$0.03\sim0.07$, instead of zero, as $L_{\rm x}$ decreases below 
$\sim 2 \times 10^{37}$ erg s$^{-1}$.

The results presented in Figure \ref{fig9} may be compared to
a theoretical prediction by Burnard et al.\ (1991),
who calculated the height of the accretion column $h_{\rm top}$ as
\begin{equation}
   \frac{h_{\rm top}}{R_{\rm NS}} \approx
   \frac{L_{\rm x}}{L_{\rm Edd}^{\rm eff} \ H_\perp} ~~ .
\label{eq5}
\end{equation}
Here, $L_{\rm Edd}^{\rm eff}$ is the Eddington luminosity along
the magnetic field which is identical to the conventional
Eddington Luminosity for a $1.4 M_\odot$ neutron star,
$L_{\rm Edd}^{\rm eff} =  2.0 \times 10^{38}$ ergs s$^{-1}$,
and $H_\perp$ is the ratio of the Thomson cross section to
the Rosseland-averaged electron scattering cross section
for radiation flows across the magnetic fields.
The dashed line in Figure \ref{fig9} shows this prediction,
assuming $H_\perp=1.23$  (Paper 1) and $R_{\rm NS}=10$ km.
Except for the observed saturation toward the highest luminosity
and $ \lesssim 2 \times 10^{37}$ erg s$^{-1}$,
the value of $h_{\rm r}$ measured at each luminosity level
thus corresponds to $\sim 70\%$ of the predicted $h_{\rm top}$.
This is quite reasonable,
because the resonance energy is expected
to sample the magnetic field strength
which is measured at, or slightly below, the top of the column
(i.e., $h_{\rm r} \lesssim h_{\rm top}$).

So far, we have assumed that the observed X-ray intensity changes 
are caused solely by actual variations of the intrinsic source luminosity.
However, the intensity may also be affected by other extrinsic factors,
such as partial obscuration by materials around the accretion disk.
Such effects are known, e.g., in Her X-1 (Mihara et al.\ 1991).
In the present outburst of 4U~0115+63, Heindl et al.\ (1999) detected
occasional appearance of ``mHz QPO'', with a period of $\sim500$ sec and
intensity changes up to $\sim40$ \%, and suggested that 
the phenomenon is possibly due to source obscuration 
by some ionized materials.
If so, we would need a caution in interpreting the results from our 
intensity-sorted analysis.

In order to investigate effects of the mHz QPO on our results,
we inspected PCA light curves acquired over the period used
for our intensity-sorted analysis (March 27 through April 8).
As shown in the inset to Figure \ref{fig10},  a relatively clear
QPO was found on March 31, 
although the period is about 1000 sec instead of the 500 sec 
reported by Heindl et al.\ (1999).
Then, we sliced the light curves as shown there, and produced
a pair of PCA spectra corresponding to peaks and valleys of the QPO. 
Figure \ref{fig10} presents these spectra and their ratio,
as well as residuals when fitted individually with the NPEX model.
The resonance energy thus clearly increases at the QPO valley.
By fitting the spectra with the NPEX times CYAB2 model
(with $W_2$ again fixed at $W_1$), we have constrained the
resonance centroid as
$10.4_{-0.7}^{+0.2}$ keV at the QPO peak
with $L_{\rm x} = 8.2 \times 10^{37}$ erg s$^{-1}$,
and
$11.8_{-0.5}^{+0.9}$ keV at the QPO bottom
with $L_{\rm x} = 5.2 \times 10^{37}$ erg s$^{-1}$.
Thus, the energy shift is statistically significant.
Furthermore, these two data points line up in Figure \ref{fig8}
closely on the general $L_{\rm x}$ vs. $E_{\rm a1}$ correlation
from the day-sorted and intensity-sorted analysis.
We hence conclude that our basic results remain unaffected,
and that the QPO, at least on this occasion, is likely to reflect
real changes in the intrinsic luminosity
rather than some obscuration effects.
This inference is reinforced by the relatively flat spectral ratio,
because the ratio should show an increased low-energy
absorption if the  obscuration were due to neutral material,
or Fe-K edge feature if it were highly ionized.

\subsection{Behavior of the other parameters}

So far, many authors (e.g. Paper 1 and Coburn et al.\ 2002)
studied relations among the continuum and cyclotron line parameters.
Our work provides a valuable opportunity to investigate
how these parameters in a single system change in correlated ways,
when the resonance energy varies.
Among various correlations,
a particularly interesting one is
that between $\tau$ and the $W/E_{\rm a}$ ratio;
in fact, Coburn et al.\ (2002) found a positive correlation
between them over a large sample of phase-averaged spectra of X-ray pulsars.
Later, Kreykenbohm et al.\ (2004) noticed
that the same correlation holds for
the phase-resolved spectra of GX~301-2,
and argued that the correlation may be explained
if the accretion column has a tall cylindrical shape
rather than flat coin-shaped geometry.
As presented in Figure \ref{fig11},
the fundamental and second harmonic widths of 4U~0115+63
from the present observations both depend positively
on their respective resonance depths,
in agreement with the correlation found by Coburn et al.\ (2002).
These results, together with the argument by Kreykenbohm et al.\ (2004),
strengthen the tall cylindrical column geometry which we invoked in \S~4.1.
For reference,
we did not find particular correlations between $E_{\rm a}$ and $kT$,
or between $E_{\rm a}$ and $W$.

Figure~\ref{fig12} shows the spectral parameters
as a function of the X-ray luminosity.
There, the positive $\tau$ vs. $W/E_{\rm a}$ correlation
of Figure~\ref{fig11} has been decomposed into
luminosity dependent changes in $W/E_{\rm a}$ (panel a)
and  $\tau$ (panels c, d).
Up to $\sim 4 \times 10^{37}$ erg s$^{-1}$,
$\tau_1$ thus stays relatively constant,
with a hint of mild increase presumably due to
the increased column density in the emission region.
Over the same luminosity range,
the fractional width $W_1/E_{\rm a1}$ increases more clearly.
This would not be due to changes in the Doppler broadening 
(e.g. M\'{e}sz\'{a}ros 1992),
since the NPEX $kT$ parameter shown in Figure~\ref{fig12}b,
which is thought to approximate the electron temperature in the
emission region (Mihara et al.\ 2004; Makishima et al.\ 1999),
stays rather constant.
Instead, the increase in the $W_1/E_{\rm a1}$ ratio could be a
result of the luminosity-correlated elongation in the accretion column,
which would cause a larger range of magnetic field intensities
to participate in the resonance formation.

Beyond $\sim 4 \times 10^{37}$ erg s$^{-1}$,
both $\tau_1$ and  $W_1/E_{\rm a1}$ start decreasing clearly,
while the 2nd harmonic resonance develops rapidly
both in depth ($\tau_2$, Figure \ref{fig12}d)
and relative width ($W_2/E_{\rm a2}$ in Figure \ref{fig12}a).
In short, the CRSF makes a transition from the single feature
at low luminosities to the harmonic double feature at higher luminosities.
Since this occurs approximately over the luminosity range
where the rapid change in $E_{\rm a1}$ takes place,
we consider that the single-to-double transition of the CRSF
has the same origin as the resonance energy shift.

The behavior of the fundamental and 2nd harmonic
parameters at the single-to-double transition may
reflect basic differences in their elementary processes.
The fundamental resonance has a very large cross section,
but it acts as scattering rather than absorption,
because an electron which is excited by absorbing a photon
of energy $\sim E_{\rm a1}$ will immediately return to the
ground state by emitting a photon of nearly the same energy.
The second harmonic resonance,
though with a much smaller cross section,
will in contrast act as pure absorption,
because the electron excitation/deexcitation in this case
occurs via absorption of a photon of energy $\sim 2 E_{\rm a1}$
and cascade emission of two photons of energies $\sim E_{\rm a1}$ each.
The emitted photons will fill up the fundamental resonance,
making it shallower
(so-called two-photon effects; Alexander \& Meszaros 1991).
In the present case,
the accretion column may be effectively thick to the fundamental
resonance ($\tau_1 > 1$) essentially at all the observed luminosities,
but presumably optically thin ($\tau_2  \ll 1$) to the second resonance photons
when the luminosity is low and hence the accretion column is short.
As the source luminosity increases,
the column becomes taller and opaque to the second resonance,
leading to the emergence of the second resonance feature.
At the same time,
the two-photon effect would reduce $\tau_1$,
just as seen in Figure \ref{fig12}c.
A similar effect was observed in the {\it INTEGRAL}
spectrum of GX~$301-2$ by Okada et al.\ (2004),
who reported that $\tau_1$ of its $\sim 35$ keV CRSF
decreased toward higher luminosities.

\subsection{The case of X0331+53}

Although 4U~0115+63 thus provides the first clear example
of luminosity-dependent changes of the cyclotron resonance energy,
there has emerged another promising case.
This is the transient X-ray pulsar X0331+53 (V0332+53),
from which a very prominent CRSF was detected with {\it Ginga}
at $E_{\rm a} = 28.5 \pm 0.5$ keV
together with a hint of the second harmonic (Makishima et al.\ 1990).
This was observed in an outburst
when the $2-60$ keV luminosity was $\sim 2 \times 10^{37}$ erg s$^{-1}$
at an assumed distance of 3 kpc.
A re-analysis of the same data using the NPEX continuum have
revised the value slightly to $E_{\rm a} = 27.2 \pm 0.3$ keV
(Makishima et al.\ 1999).

\begin{figure}
\plotone{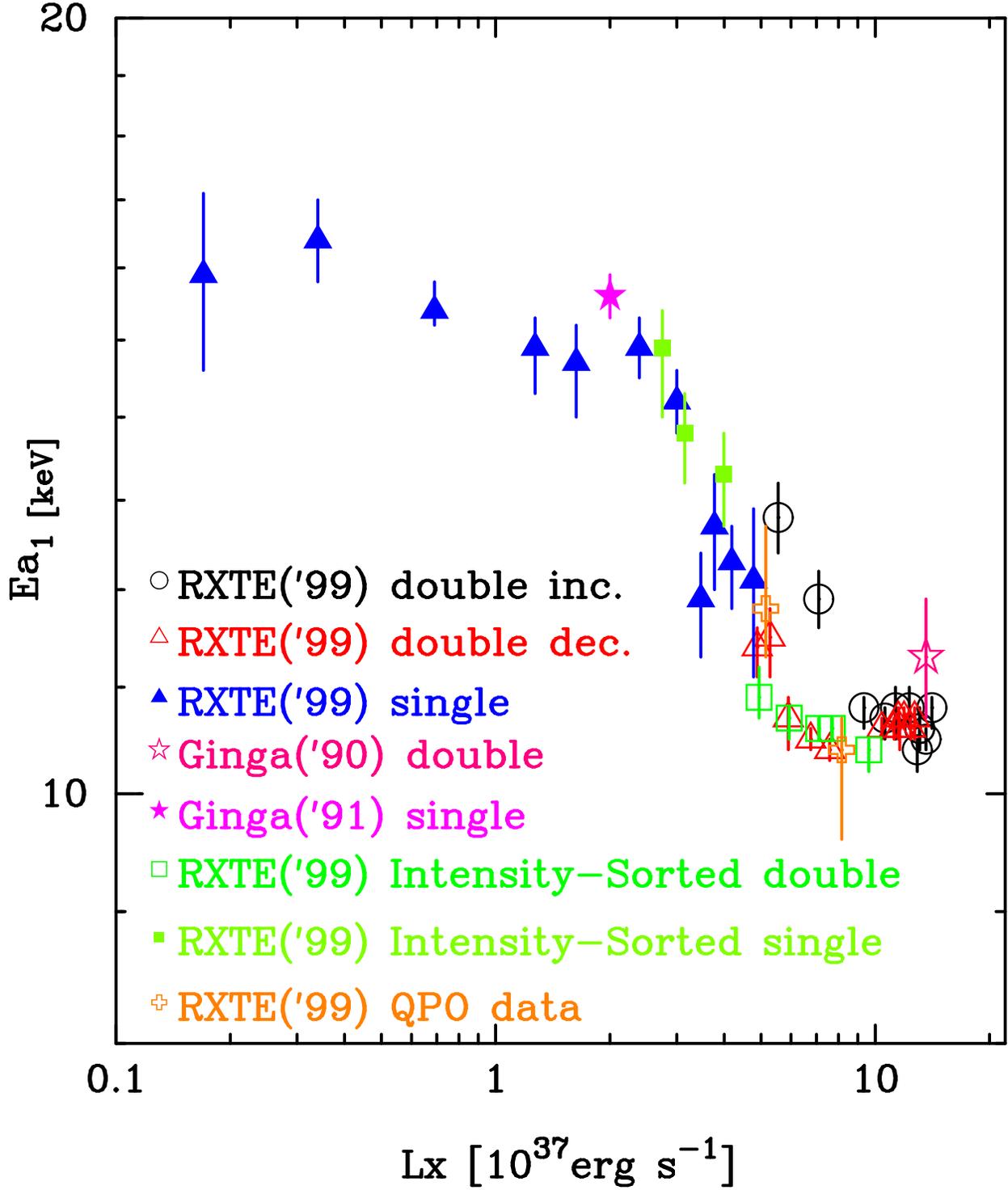}
\caption{The obtained fundamental cyclotron energies shown against the $3-50$ keV luminosity. The open symbols represent the fundamental energy of the double absorption features, and the filled symbols the energy of the single CRSF. All the date-sorted data in the brightening (circles) and declining (triangles) are presented together with the intensity-sorted (squares) results. The {\it Ginga} results (Paper 1) are also included with asterisks. }
\label{fig8}
\end{figure}

\begin{figure}
\plotone{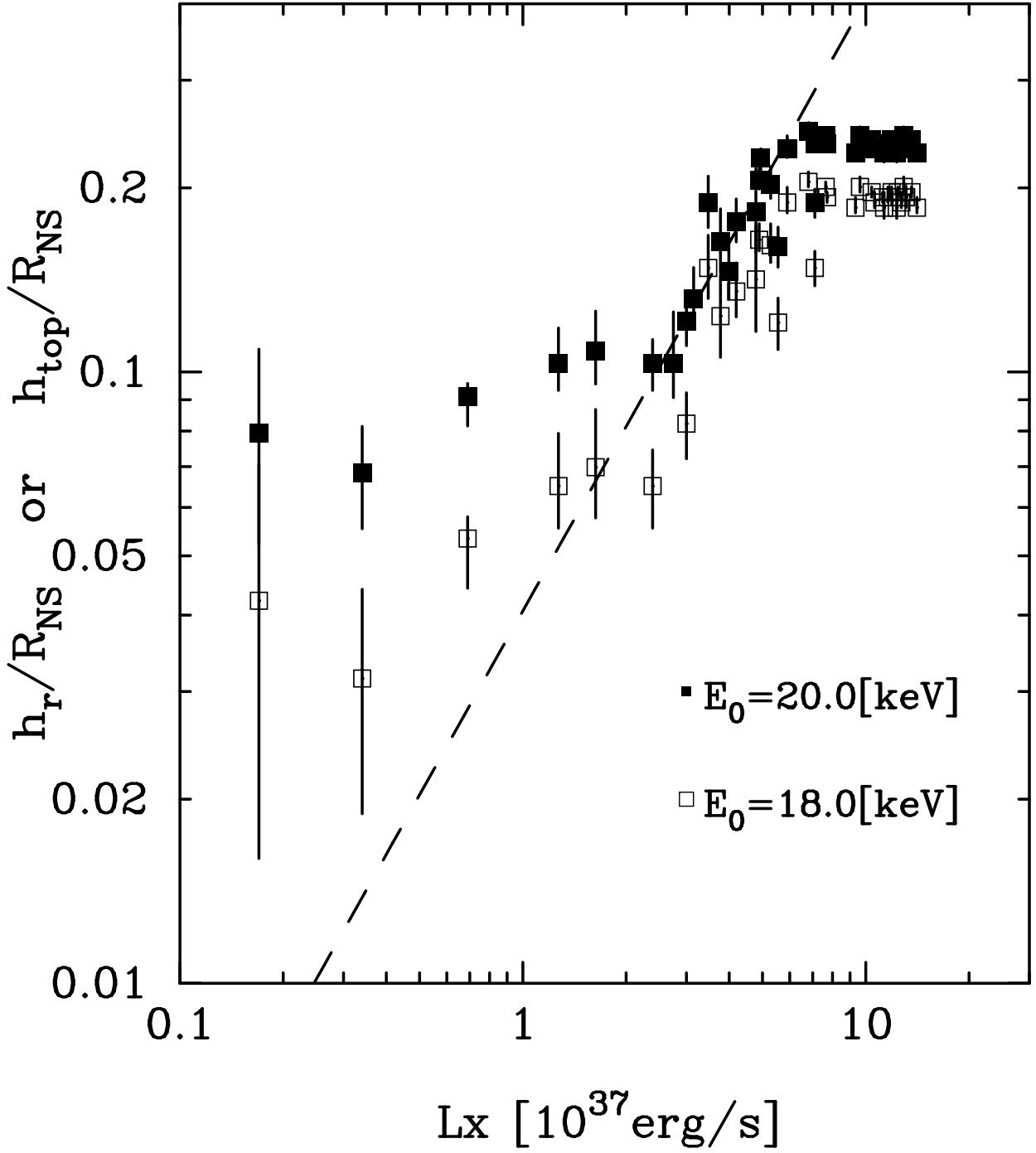}
\caption{The same as Figure \ref{fig8}, but the resonance energy is converted through equation \ref{eq4} into the height ($h_{\rm r}$) at which the resonance occurs. The dashed line represents the value of the column top height, $h_{\rm top}$, calculated by equation \ref{eq5}. }
\label{fig9}
\end{figure}

\begin{figure}
\epsscale{0.7}
\plotone{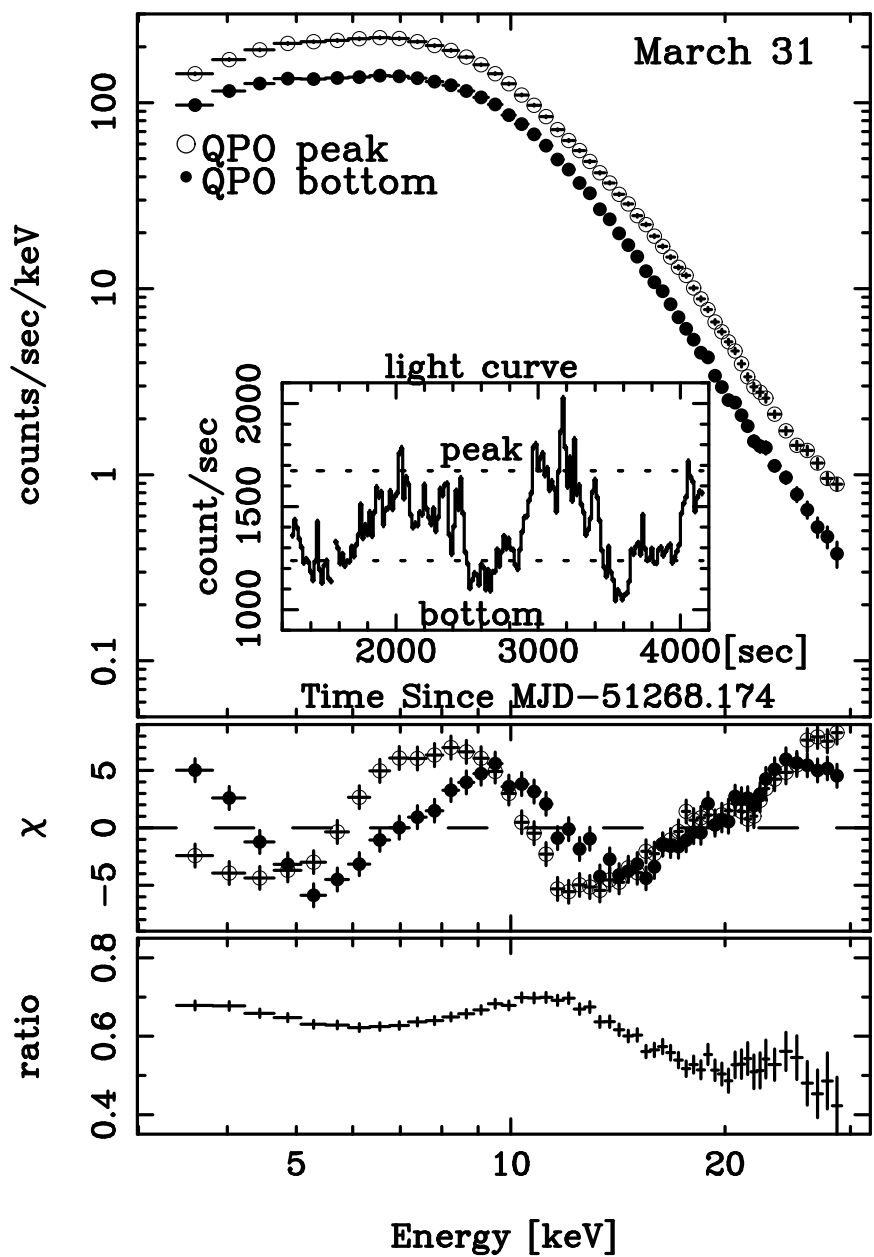}
\caption{Spectral changes associated with a 1 mHz QPO observed on March 31.  The top panel shows the background-subtracted PCA spectra accumulated at peaks and bottoms of the QPO, as specified on the $2-30$ keV light curve (inset).  The middle panel is the residuals from NPEX fits to the two spectra.  The bottom pannel show the QPO bottom-to-peak ratio. }
\label{fig10}
\end{figure}

\begin{figure}
\epsscale{0.7}
\plotone{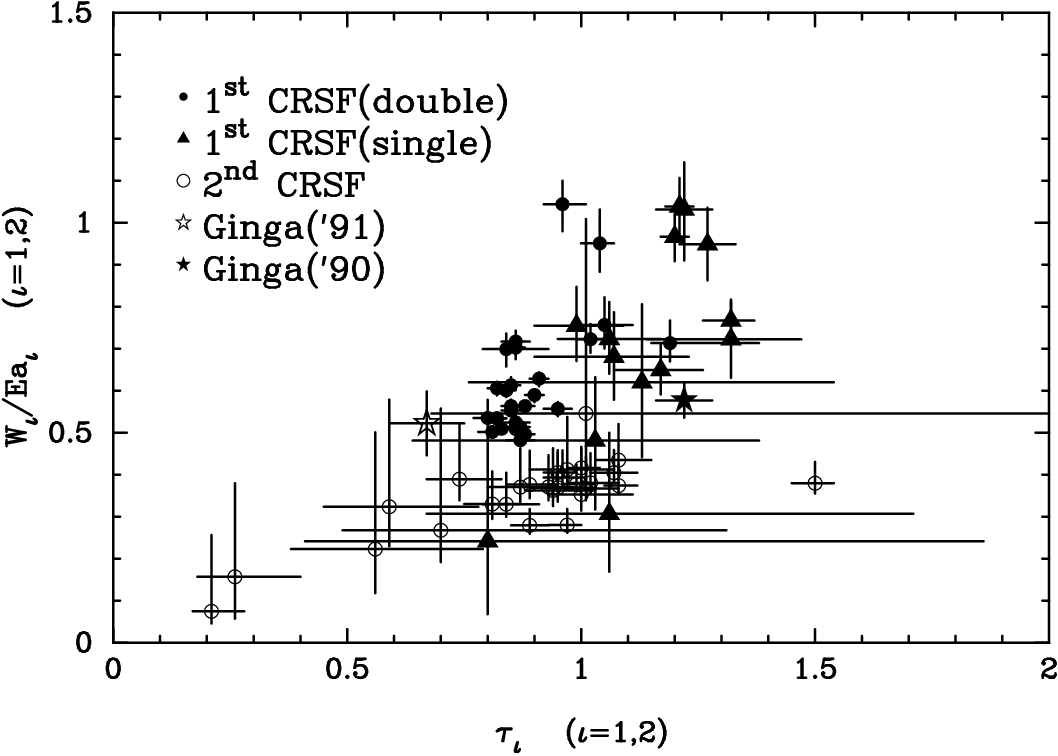}
\caption{The CRSF depth $\tau$ against the fractional width $W/E_{\rm a}$. The plotted points are obtained from the date-sorted and intensity-sorted analyses. }
\label{fig11}
\end{figure}

\begin{figure}
\epsscale{1.0}
\plotone{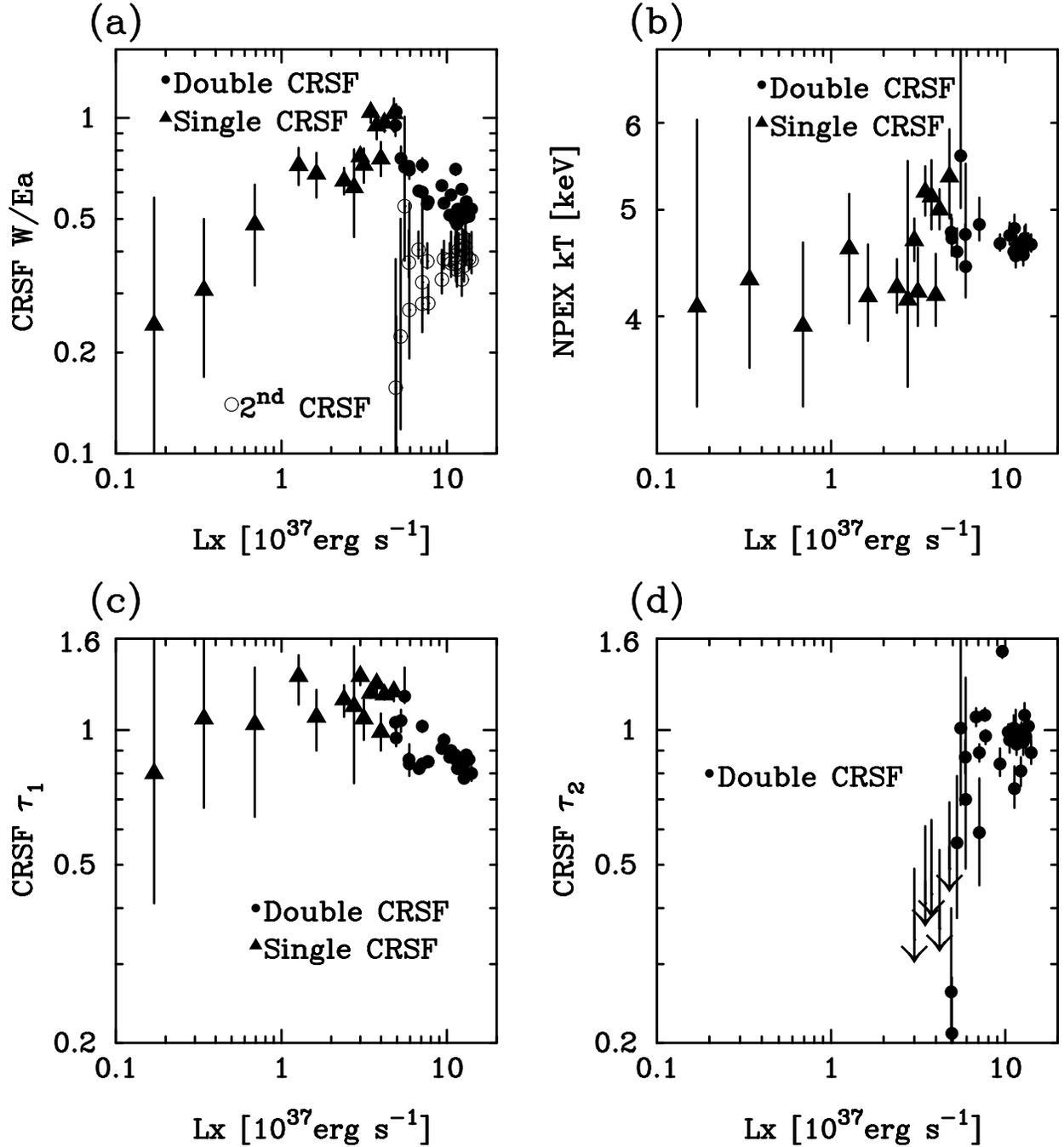}
\caption{The dependence of the CRSF and NPEX parameters on the $3-50$ keV luminosity, obtained from the 34 date-sorted data and the 8 intensity-sorted data. The circles and triangles represent those spectra which are described by the double and single CRSF factor(s), respectively. (a) The $W/Ea$ ratio for the fundamental and second harmonic resonances. (b) The NPEX $kT$. (c) The fundamental resonance depth, $\tau_1$. (d) The second harmonic depth, $\tau_2$. }
\label{fig12}
\end{figure}

The object entered a bright outburst from 2004 November,
becoming considerably more luminous than was observed
with {\it Ginga} (Swank et al.\ 2004; Remillard 2004).
Observations with {\it INTEGRAL} (Kreykenbohm et al.\ 2005) and {\it RXTE}
(Coburn et al.\ 2004) have clearly reconfirmed the fundamental CRSF,
and further revealed the second and third resonances.
The fundamental resonance energy obtained with {\it INTEGRAL}
is $24.9 \pm 0.1$ keV (Kreykenbohm et al. 2005)
at a $3-50$ keV luminosity of $\sim8\times10^{37}$ erg s$^{-1}$.
This means that a factor 4 increase in the luminosity
(from the {\it Ginga} outburst to the present one)
is accompanied by a $\sim 10\%$ decrease in $E_{\rm a1}$.
As mentioned by Mihara et al.\ (2004),
the same effect was already visible between
two pointings with {\it Ginga}.
For comparison,
the resonance energy of 4U~0115+63 changes
by a somewhat larger amount ($\sim 20\%$)
across the same factor 4 luminosity range (Figure \ref{fig8}).
The archival {\it RXTE} data of X0331+53 are currently
being analyzed,
and the results will be reported elsewhere
(Nakajima et al.\ , in preparation).

Now, we have two examples of luminosity
dependent changes in the resonance energy.
However, there is an apparent counter example,
namely Her X-1.
In a large amount of {\it RXTE} data covering its ``main on'' phase,
the $2-30$ keV luminosity (at 5.8 kpc)
of Her X-1 varied by a factor of 2
over  $(1.8-3.2) \times 10^{37}$ erg $s^{-1}$,
but the CRSF stayed at $\sim 40$ keV (Gruber et al.\ 2001).
One possible interpretation of these results on Her X-1 is
that its luminosity swing was not large enough
to reach the critical range ({\it cf.} Figure~\ref{fig8})
where $E_{\rm a}$ starts changing significantly.
Alternatively,
the luminosity-related changes
in the CRSF may depend on the object,
for some reasons which are yet to be detailed.

In conclusion,
our study using the {\it RXTE} data
has confirmed the inference made in Paper 1,
that the cyclotron resonance energy of 4U~0115+63
increases as the X-ray luminosity decreases.
While this provide a new tool with which we can diagnose
the accretion column of strongly magnetized neutron stars,
it remains yet to be confirmed
whether the phenomenon is common among this type of objects.

\clearpage

\begin{deluxetable}{ccccccccc}
\tabletypesize{\scriptsize} 
\tablecaption{The log of {\it RXTE} observations of 4U~0115+63 in the 1999 March$-$April outburst. \label{tab1}}
\tablewidth{0pt}
\tablehead{
 \colhead{} & \colhead{} & \colhead{} &
 \multicolumn{3}{c}{PCA} & 
 \colhead{} &
 \multicolumn{2}{c}{HEXTE} \\
 \cline{4-6} \cline{8-9} \\
 \colhead{Obs.} & 
 \colhead{Date} &  
 \colhead{Start/End Time\tablenotemark{a}} &
 \colhead{PCU. } &
 \colhead{Exposure} &
 \colhead{Rate\tablenotemark{b}} &
 \colhead{} &
 \colhead{Exposure} &
 \colhead{Rate\tablenotemark{c}} \\
 \colhead{} &
 \colhead{(1999)} &
 \colhead{(UT)} &
 \colhead{No.} &
 \colhead{[ks]} &
 \colhead{[c s$^{-1}$ PCU$^{-1}$]} &
 \colhead{} & 
 \colhead{[ks]} &
 \colhead{[c s$^{-1}$]}
}
\startdata
1  & Mar  3 & 03:35/03:49 & all     & 0.57 &  556 $\pm$ 1.1 & & 0.18 & 48.8 $\pm$ 0.9 \\
2  & Mar  4 & 13:08/13:41 & 0,1,2   & 1.01 &  466 $\pm$ 1.0 & & 0.15 & 37.2 $\pm$ 0.8 \\
3  & Mar  5 & 20:35/20:57 & all     & 1.28 &  713 $\pm$ 1.4 & & 0.44 & 70.3 $\pm$ 0.6 \\
4  & Mar  6 & 20:47/20:56 & all     & 0.54 &  806 $\pm$ 1.6 & & 0.18 & 82.8 $\pm$ 1.0 \\
5  & Mar  7 & 20:40/20:55 & all     & 0.84 &  866 $\pm$ 1.7 & & 0.29 & 85.3 $\pm$ 0.8 \\
6  & Mar  9 & 06:22/06:36 & 0,2,3,4 & 0.81 &  918 $\pm$ 1.8 & & 0.20 & 93.3 $\pm$ 0.9 \\
7  & Mar 11 & 08:41/09:40 & 0,2     & 1.60 & 1020 $\pm$ 2.0 & & 0.40 & 97.7 $\pm$ 0.6 \\
8  & Mar 13 & 18:58/19:41 & 0,2,3,4 & 0.67 & 1011 $\pm$ 2.0 & & 0.21 & 104  $\pm$ 1.0 \\ 
9  & Mar 14 & 06:44/06:56 & 0,2,3   & 0.68 & 1055 $\pm$ 2.1 & & 0.27 & 105  $\pm$ 1.0 \\ 
10 & Mar 15 & 07:53/08:17 & 0,2     & 0.79 &  994 $\pm$ 2.1 & & 0.28 & 101  $\pm$ 1.0 \\ 
11 & Mar 16 & 09:33/09:46 & 0,2     & 0.25 &  997 $\pm$ 2.4 & & 0.09 & 99.9 $\pm$ 1.4 \\ 
12 & Mar 18 & 07:49/08:05 & 0,2,3,4 & 0.37 &  949 $\pm$ 2.0 & & 0.13 & 98.5 $\pm$ 1.2 \\ 
13 & Mar 19a& 04:26/06:13 & 0,2,3   & 3.45 &  924 $\pm$ 1.8 & & 1.07 & 92.7 $\pm$ 0.4 \\ 
14 & Mar 19b& 12:51/13:12 & 0,2     & 0.33 &  876 $\pm$ 2.1 & & 0.10 & 96.0 $\pm$ 1.5 \\ 
15 & Mar 20 & 03:34/03:57 & 0,2     & 0.88 &  857 $\pm$ 1.8 & & 0.22 & 85.2 $\pm$ 0.9 \\ 
16 & Mar 21a& 06:27/08:33 & 0,2,3,4 & 4.13 &  875 $\pm$ 1.7 & & 1.46 & 91.2 $\pm$ 0.3 \\ 
17 & Mar 21b& 12:48/13:27 & 0,2,3,4 & 0.93 &  852 $\pm$ 1.7 & & 0.29 & 88.9 $\pm$ 0.9 \\ 
18 & Mar 22 & 04:22/05:28 & 0,2     & 3.11 &  788 $\pm$ 1.6 & & 1.06 & 82.0 $\pm$ 0.4 \\ 
19 & Mar 27 & 07:53/11:01 & all     & 4.02 &  564 $\pm$ 1.2 & & 1.42 & 60.2 $\pm$ 0.4 \\ 
20 & Mar 29 & 00:12/03:01 & 0,1,2,3 & 4.26 &  517 $\pm$ 1.1 & & 1.40 & 49.2 $\pm$ 0.3 \\ 
21 & Mar 31 & 04:11/06:24 & 0,2,3   & 4.11 &  456 $\pm$ 1.0 & & 1.41 & 41.5 $\pm$ 0.3 \\ 
22 & Apr  1 & 03:27/04:19 & 0,1,2   & 0.27 &  437 $\pm$ 1.2 & & 0.09 & 37.8 $\pm$ 1.1 \\ 
23 & Apr  2 & 04:09/06:14 & all     & 3.63 &  399 $\pm$ 0.8 & & 1.16 & 32.7 $\pm$ 0.4 \\ 
24 & Apr  3 & 02:30/02:45 & 0,2,3   & 0.50 &  385 $\pm$ 0.9 & & 0.16 & 30.8 $\pm$ 1.0 \\ 
25 & Apr  4 & 04:08/06:11 & all     & 3.47 &  345 $\pm$ 0.7 & & 1.16 & 26.9 $\pm$ 0.4 \\ 
26 & Apr  5 & 11:09/11:27 & 0,2,3   & 0.89 &  309 $\pm$ 0.7 & & 0.33 & 23.0 $\pm$ 0.6 \\ 
27 & Apr  6 & 02:25/04:28 & all     & 3.80 &  287 $\pm$ 0.6 & & 1.25 & 20.8 $\pm$ 0.3 \\ 
28 & Apr  8 & 01:13/03:27 & 0,2,3,4 & 4.66 &  254 $\pm$ 0.6 & & 1.56 & 17.3 $\pm$ 0.3 \\ 
29 & Apr 10 & 03:59/06:05 & all     & 3.48 &  210 $\pm$ 0.5 & & 1.18 & 12.1 $\pm$ 0.3 \\ 
30 & Apr 12 & 02:16/04:28 & 0,1,2,3 & 4.12 &  146 $\pm$ 0.3 & & 1.36 &  8.6 $\pm$ 0.3 \\ 
31 & Apr 14 & 01:14/03:22 & all     & 3.94 &  111 $\pm$ 0.2 & & 1.31 &  6.8 $\pm$ 0.3 \\ 
32 & Apr 16 & 21:58/22:12 & 0,1,2,4 & 0.83 & 60.1 $\pm$ 0.2 & & 0.28 &  2.9 $\pm$ 0.6 \\ 
33 & Apr 18 & 21:56/22:10 & 0,2,3,4 & 0.82 & 24.9 $\pm$ 0.1 & & 0.28 &  1.9 $\pm$ 0.6 \\ 
34 & Apr 20 & 01:14/01:34 & all     & 0.64 & 12.4 $\pm$ 0.1 & & 0.22 &  0.7 $\pm$ 0.6 \\ 
\enddata

\tablenotetext{a}{
Start and end time (UT) of the PCA observations. 
}
\tablenotetext{b}{
In the 3-30 keV energy range.
}
\tablenotetext{c}{
Count rates of HEXTE cluster A in the 15-50 keV energy range.
}
\end{deluxetable}

\clearpage

\begin{deluxetable}{cccccccccccccccccccccccc}
\tabletypesize{\tiny}
\rotate
\tablecaption{Summary of the best-fit parameters of the NPEX$\times$CYAB($\times$CYAB) model, determined by the date-sorted PCA spectra from March 27 thorough April 8.
\label{tab2}}
\tablewidth{0pt}
\tablehead{
 \colhead{Date} &
 \colhead{$A_1$} &
 \colhead{$\alpha_1$} & 
 \colhead{$A_2^{a}$} &
 \colhead{$kT$} &
 \colhead{$E_{a1}^{b}$} &
 \colhead{$W_1$} &
 \colhead{$\tau_1$} &
 \colhead{$W_2$} &
 \colhead{$\tau_2$} &
 \colhead{$L_{\rm x}^{c}$} &
 \colhead{$\chi^2_{\nu}$} \\
 \colhead{  } &
 \colhead{(ph s$^{-1}$ keV$^{-1}$)} &
 \colhead{  } & 
 \colhead{(ph s$^{-1}$ keV$^{-1}$)} &
 \colhead{(keV)} &
 \colhead{(keV)} &
 \colhead{(keV)} &
 \colhead{  } &
 \colhead{(keV)} &
 \colhead{  } &
 \colhead{  } &
 \colhead{  } &
}
\startdata
Mar 27 &$ 0.50_{-0.22}^{+0.45} $&$ 1.49_{-0.49}^{+0.53} $&$ 7.92_{-1.83}^{+1.15} $&$ 4.91_{-0.43}^{+1.12} $&$ 10.6_{-0.1}^{+0.1} $&$ 5.91_{-0.15}^{+0.17} $&$ 0.89_{-0.05}^{+0.05} $&$ 6.80_{-1.29}^{+1.93} $&$ 1.12_{-0.25}^{+0.51} $& 6.79 & 0.62 \\
Mar 29 &$ 0.57_{-0.25}^{+0.50} $&$ 1.67_{-0.49}^{+0.53} $&$ 7.88_{-1.24}^{+0.82} $&$ 4.61_{-0.28}^{+0.61} $&$ 10.6_{-0.1}^{+0.1} $&$ 6.59_{-0.25}^{+0.27} $&$ 0.85_{-0.04}^{+0.06} $&$ 5.91_{-1.18}^{+1.67} $&$ 0.91_{-0.19}^{+0.34} $& 6.04 & 0.89 \\
Mar 31 &$ 0.53_{-0.20}^{+0.34} $&$ 1.68_{-0.38}^{+0.40} $&$ 6.10_{-0.22}^{+0.21} $&$ 5.22 {\rm (fixed)}   $&$ 10.7_{-0.2}^{+0.2} $&$ 7.78_{-0.26}^{+0.29} $&$ 0.95_{-0.04}^{+0.03} $&$ 7.47_{-0.67}^{+0.69} $&$ 1.03_{-0.08}^{+0.07} $& 5.25 & 0.75 \\
Apr  2 &$ 0.61_{-0.22}^{+0.25} $&$ 1.84_{-0.37}^{+0.28} $&$ 5.26_{-0.18}^{+0.13} $&$ 5.39 {\rm (fixed)}   $&$ 11.5_{-0.3}^{+0.3} $&$ 10.25_{-0.49}^{+0.76} $&$ 1.12_{-0.06}^{+0.05} $&$ 6.33_{-1.93}^{+1.65} $&$ 0.59_{-0.20}^{+0.15} $& 4.61 & 1.13 \\
Apr  3 &$ 0.61_{-0.27}^{+0.55} $&$ 1.79_{-0.47}^{+0.53} $&$ 6.59_{-1.10}^{+0.90} $&$ 3.88_{-0.33}^{+0.70} $&$ 13.6_{-1.0}^{+0.5} $&$ 8.94_{-1.63}^{+2.27} $&$ 0.88_{-0.17}^{+0.22} $&$ - $&$ - $& 4.24 & 0.90 \\
Apr  4 &$ 0.57_{-0.21}^{+0.37} $&$ 1.85_{-0.38}^{+0.41} $&$ 5.80_{-0.49}^{+0.45} $&$ 3.95_{-0.22}^{+0.32} $&$ 13.8_{-0.5}^{+0.4} $&$ 9.05_{-0.98}^{+1.16} $&$ 0.93_{-0.10}^{+0.11} $&$ - $&$ - $& 3.85 & 0.93 \\
Apr  5 &$ 0.31_{-0.11}^{+0.22} $&$ 1.46_{-0.38}^{+0.42} $&$ 3.93_{-1.01}^{+0.88} $&$ 4.96_{-0.83}^{+2.59} $&$ 13.2_{-1.6}^{+0.8} $&$ 11.28_{-2.02}^{+2.89} $&$ 1.27_{-0.24}^{+0.26} $&$ - $&$ - $& 3.42 & 0.60 \\ 
Apr  6 &$ 0.43_{-0.16}^{+0.28} $&$ 1.73_{-0.37}^{+0.40} $&$ 4.89_{-0.42}^{+0.40} $&$ 3.92_{-0.22}^{+0.32} $&$ 13.7_{-0.5}^{+0.4} $&$ 8.98_{-1.04}^{+1.19} $&$ 0.91_{-0.10}^{+0.11} $&$ - $&$ - $& 3.19 & 0.97 \\
Apr  8 &$ 0.19_{-0.06}^{+0.10} $&$ 1.14_{-0.30}^{+0.35} $&$ 3.92_{-0.48}^{+0.52} $&$ 3.97_{-0.29}^{+0.38} $&$ 15.0_{-0.4}^{+0.3} $&$ 8.68_{-1.14}^{+1.18} $&$ 1.06_{-0.15}^{+0.16} $&$ - $&$ - $& 2.78 & 0.94 \\
\enddata

\tablenotetext{a}{$\times 10^{-3}$}
\tablenotetext{b}{The second resonance energy $Ea2$ was fixed at $2 \times E_{a1}$.}
\tablenotetext{c}{$10^{37}$ erg s$^{-1}$ in 3$-$30 keV}

\end{deluxetable}

\clearpage

\begin{deluxetable}{cccccccccccccccccccccccc}
\tabletypesize{\tiny}
\rotate
\tablecaption{The same as Table \ref{tab2}, but for the intensity-sorted PCA spectra.
\label{tab3}}
\tablewidth{0pt}
\tablehead{
 \colhead{Region} &
 \colhead{$A_1$} &
 \colhead{$\alpha_1$} & 
 \colhead{$A_2^{a}$} &
 \colhead{$kT$} &
 \colhead{$E_{a1}^{b}$} &
 \colhead{$W_1$} &
 \colhead{$\tau_1$} &
 \colhead{$W_2$} &
 \colhead{$\tau_2$} &
 \colhead{$L_{\rm x}^{c}$} &
 \colhead{$\chi^2_{\nu}$} \\
 \colhead{  } &
 \colhead{(ph s$^{-1}$ keV$^{-1}$)} &
 \colhead{  } & 
 \colhead{(ph s$^{-1}$ keV$^{-1}$)} &
 \colhead{(keV)} &
 \colhead{(keV)} &
 \colhead{(keV)} &
 \colhead{  } &
 \colhead{(keV)} &
 \colhead{  } &
 \colhead{  } &
 \colhead{  } &
}
\startdata
f1 &$ 0.42_{-0.17}^{+0.30} $&$ 1.27_{-0.43}^{+0.45} $&$ 7.20_{-0.46}^{+0.39} $&$ 5.76 {\rm (fixed)}   $&$ 10.4_{-0.2}^{+0.1} $&$ 5.79_{-0.14}^{+0.16} $&$ 0.95_{-0.03}^{+0.03} $&$ 7.90_{-0.49}^{+0.52} $&$ 1.50_{-0.05}^{+0.04} $& 7.47 & 0.84 \\
f2 &$ 0.58_{-0.24}^{+0.45} $&$ 1.61_{-0.46}^{+0.47} $&$ 8.76_{-0.34}^{+0.28} $&$ 4.59 {\rm (fixed)}   $&$ 10.6_{-0.1}^{+0.1} $&$ 5.97_{-0.14}^{+0.16} $&$ 0.85_{-0.02}^{+0.02} $&$ 5.94_{-0.39}^{+0.40} $&$ 0.97_{-0.04}^{+0.03} $& 6.81 & 0.72 \\
f3 &$ 0.62_{-0.27}^{+0.49} $&$ 1.73_{-0.47}^{+0.48} $&$ 8.46_{-0.30}^{+0.25} $&$ 4.54 {\rm (fixed)}   $&$ 10.6_{-0.1}^{+0.1} $&$ 6.37_{-0.16}^{+0.19} $&$ 0.84_{-0.02}^{+0.02} $&$ 5.92_{-0.42}^{+0.41} $&$ 0.89_{-0.04}^{+0.04} $& 6.33 & 0.68 \\
f4 &$ 0.83_{-0.43}^{+0.86} $&$ 2.05_{-0.61}^{+0.59} $&$ 7.69_{-1.93}^{+0.84} $&$ 4.44_{-0.28}^{+1.07} $&$ 10.7_{-0.2}^{+0.1} $&$ 7.48_{-0.42}^{+0.39} $&$ 0.84_{-0.05}^{+0.09} $&$ 5.73_{-1.62}^{+3.10} $&$ 0.70_{-0.21}^{+0.61} $& 5.39 & 0.64 \\
f5 &$ 0.99_{-0.41}^{+0.76} $&$ 2.26_{-0.43}^{+0.46} $&$ 6.48_{-0.40}^{+0.38} $&$ 4.71_{-0.17}^{+0.20} $&$ 10.9_{-0.2}^{+0.3} $&$ 11.38_{-0.67}^{+0.52} $&$ 0.96_{-0.04}^{+0.05} $&$ 1.63_{-0.63}^{+1.98} $&$ 0.21_{-0.04}^{+0.07} $& 4.59 & 0.91 \\
f6 &$ 0.56_{-0.20}^{+0.34} $&$ 1.84_{-0.36}^{+0.39} $&$ 5.43_{-0.45}^{+0.42} $&$ 4.18_{-0.26}^{+0.38} $&$ 13.3_{-0.6}^{+0.5} $&$ 10.04_{-1.02}^{+1.17} $&$ 0.99_{-0.09}^{+0.10} $&$ -                    $&$ -                    $& 3.79 & 0.97 \\
f7 &$ 0.29_{-0.10}^{+0.16} $&$ 1.45_{-0.34}^{+0.36} $&$ 4.17_{-0.39}^{+0.39} $&$ 4.21_{-0.29}^{+0.41} $&$ 13.8_{-0.6}^{+0.5} $&$ 9.98_{-1.06}^{+1.16} $&$ 1.06_{-0.11}^{+0.11} $&$  -                    $&$ -                    $& 2.99 & 0.86 \\
f8 &$ 0.15_{-0.05}^{+0.16} $&$ 0.99_{-0.38}^{+0.60} $&$ 3.48_{-1.03}^{+1.28} $&$ 4.14_{-0.69}^{+1.40} $&$ 14.9_{-0.9}^{+0.5} $&$ 9.24_{-2.60}^{+2.75} $&$ 1.13_{-0.37}^{+0.41} $&$  -                    $&$ -                    $& 2.62 & 0.73 \\
\enddata

\tablenotetext{a}{$\times 10^{-3}$}
\tablenotetext{b}{The second resonance energy was fixed at $2 \times E_{a1}$.}
\tablenotetext{c}{$10^{37}$ erg s$^{-1}$ in 3$-$30 keV}

\end{deluxetable}

\clearpage

\begin{deluxetable}{cccccccccccccccccccccccc}
\tabletypesize{\tiny}
\rotate
\tablecaption{The best-fit parameters of the date-sorted spectra, incorporating the HEXTE data.
\label{tab4}}
\tablewidth{0pt}
\tablehead{
 \colhead{Observation} &
 \colhead{$A_1$} &
 \colhead{$\alpha_1$} & 
 \colhead{$A_2^{a}$} &
 \colhead{$kT$} &
 \colhead{$E_{a1}^{b}$} &
 \colhead{$W_1$} &
 \colhead{$\tau_1$} &
 \colhead{$W_2$} &
 \colhead{$\tau_2$} &
 \colhead{$L_{\rm x}^{c}$} &
 \colhead{$\chi^2_{\nu}$} \\
 \colhead{(1999)} &
 \colhead{(ph s$^{-1}$ keV$^{-1}$)} &
 \colhead{  } & 
 \colhead{(ph s$^{-1}$ keV$^{-1}$)} &
 \colhead{(keV)} &
 \colhead{(keV)} &
 \colhead{(keV)} &
 \colhead{  } &
 \colhead{(keV)} &
 \colhead{  } &
 \colhead{  } &
 \colhead{  } &
}
\startdata
Mar  3  &$ 0.40_{-0.13}^{+0.21} $&$ 1.24_{-0.35}^{+0.36} $&$ 7.47_{-0.68}^{+0.56}  $&$ 4.85_{-0.16}^{+0.28} $&$ 11.9\pm0.3         $&$ 8.60_{-0.32}^{+0.37} $&$ 1.02\pm0.03          $&$ 7.70_{-2.22}^{+3.03}  $&$ 0.59_{-0.14}^{+0.19} $& 7.10 & 0.97 \\
Mar  4  &$ 0.36_{-0.10}^{+0.17} $&$ 1.20_{-0.27}^{+0.32} $&$ 4.80_{-1.76}^{+0.80}  $&$ 5.60_{-0.58}^{+6.90} $&$ 12.8\pm0.4         $&$ 9.13_{-0.48}^{+0.63} $&$ 1.19_{-0.04}^{+0.19} $&$ 13.97_{-4.34}^{+5.91} $&$ 1.01_{-0.33}^{+1.27} $& 5.55 & 1.33 \\
Mar  5  &$ 0.56_{-0.20}^{+0.35} $&$ 1.37_{-0.37}^{+0.41} $&$ 10.87_{-0.50}^{+0.46} $&$ 4.66_{-0.07}^{+0.09} $&$ 10.8_{-0.2}^{+0.1} $&$ 6.79\pm0.15          $&$ 0.91\pm0.02          $&$ 7.12_{-0.63}^{+0.81}  $&$ 0.84_{-0.05}^{+0.07} $& 9.33 & 1.37 \\
Mar  6  &$ 0.59_{-0.21}^{+0.39} $&$ 1.29_{-0.39}^{+0.44} $&$ 12.07_{-0.70}^{+0.61} $&$ 4.74_{-0.09}^{+0.13} $&$ 10.7_{-0.2}^{+0.1} $&$ 6.31_{-0.14}^{+0.13} $&$ 0.90\pm0.02          $&$ 7.90_{-0.70}^{+0.95}  $&$ 0.95_{-0.06}^{+0.08} $& 10.6 & 1.24 \\
Mar  7  &$ 0.42_{-0.14}^{+0.25} $&$ 0.89_{-0.35}^{+0.40} $&$ 11.84_{-0.88}^{+0.79} $&$ 4.81_{-0.10}^{+0.14} $&$ 10.8_{-0.3}^{+0.2} $&$ 7.59_{-0.21}^{+0.22} $&$ 0.86_{-0.03}^{+0.02} $&$ 8.41_{-1.06}^{+1.41}  $&$ 0.74_{-0.07}^{+0.09} $& 11.3 & 1.28 \\
Mar  9  &$ 0.51_{-0.17}^{+0.28} $&$ 0.96_{-0.36}^{+0.39} $&$ 13.05_{-0.92}^{+0.79} $&$ 4.67\pm0.08          $&$ 10.8\pm0.2         $&$ 6.62_{-0.16}^{+0.17} $&$ 0.85_{-0.03}^{+0.02} $&$ 7.13_{-0.75}^{+0.83}  $&$ 0.81_{-0.06}^{+0.06} $& 12.3 & 0.85 \\
Mar 11  &$ 0.70_{-0.23}^{+0.41} $&$ 1.14_{-0.34}^{+0.40} $&$ 14.75_{-0.84}^{+0.78} $&$ 4.69_{-0.06}^{+0.08} $&$ 10.6_{-0.2}^{+0.1} $&$ 5.97_{-0.10}^{+0.11} $&$ 0.88\pm0.02          $&$ 8.34_{-0.52}^{+0.68}  $&$ 0.96_{-0.04}^{+0.05} $& 13.1 & 1.18 \\
Mar 13  &$ 0.69_{-0.24}^{+0.46} $&$ 1.17_{-0.38}^{+0.44} $&$ 15.00_{-0.93}^{+0.82} $&$ 4.66_{-0.08}^{+0.10} $&$ 10.5\pm0.1         $&$ 5.34\pm0.09          $&$ 0.86_{-0.03}^{+0.02} $&$ 7.99_{-0.55}^{+0.74}  $&$ 1.02_{-0.05}^{+0.06} $& 13.6 & 1.37 \\
Mar 14  &$ 0.41_{-0.13}^{+0.22} $&$ 0.70_{-0.35}^{+0.39} $&$ 14.27_{-1.19}^{+1.02} $&$ 4.65_{-0.07}^{+0.10} $&$ 10.8_{-0.2}^{+0.1} $&$ 5.78_{-0.11}^{+0.12} $&$ 0.80\pm0.03          $&$ 8.13_{-0.68}^{+0.87}  $&$ 0.89_{-0.05}^{+0.06} $& 14.1 & 1.23 \\
Mar 15  &$ 0.88_{-0.32}^{+0.67} $&$ 1.40_{-0.40}^{+0.48} $&$ 14.88_{-0.83}^{+0.76} $&$ 4.71_{-0.08}^{+0.12} $&$ 10.4_{-0.2}^{+0.1} $&$ 5.34_{-0.09}^{+0.11} $&$ 0.86\pm0.02          $&$ 9.04_{-0.58}^{+0.89}  $&$ 1.08_{-0.05}^{+0.07} $& 12.9 & 1.56 \\
Mar 16  &$ 0.80_{-0.34}^{+0.70} $&$ 1.31_{-0.48}^{+0.54} $&$ 14.66_{-1.19}^{+1.01} $&$ 4.67_{-0.13}^{+0.18} $&$ 10.6\pm0.2         $&$ 5.56_{-0.15}^{+0.17} $&$ 0.86\pm0.03          $&$ 8.74_{-1.00}^{+1.32}  $&$ 0.97_{-0.08}^{+0.10} $& 13.1 & 1.21 \\
Mar 18  &$ 0.55_{-0.20}^{+0.37} $&$ 1.03_{-0.39}^{+0.44} $&$ 13.95_{-1.04}^{+0.90} $&$ 4.55_{-0.10}^{+0.14} $&$ 10.7_{-0.2}^{+0.1} $&$ 5.37\pm0.11          $&$ 0.81\pm0.03          $&$ 7.76_{-0.78}^{+1.07}  $&$ 0.94_{-0.06}^{+0.09} $& 12.7 & 1.47 \\
Mar 19a &$ 0.56_{-0.17}^{+0.34} $&$ 1.10_{-0.31}^{+0.41} $&$ 13.29_{-0.65}^{+0.68} $&$ 4.69_{-0.04}^{+0.06} $&$ 10.5_{-0.1}^{+0.1} $&$ 5.35_{-0.07}^{+0.08} $&$ 0.83_{-0.02}^{+0.02} $&$ 8.74_{-0.34}^{+0.52}  $&$ 1.00_{-0.03}^{+0.04} $& 12.4 & 2.16 \\
Mar 19b &$ 0.70_{-0.28}^{+0.51} $&$ 1.30_{-0.43}^{+0.47} $&$ 13.07_{-1.00}^{+0.80} $&$ 4.54_{-0.11}^{+0.17} $&$ 10.7_{-0.2}^{+0.1} $&$ 5.16_{-0.13}^{+0.13} $&$ 0.87_{-0.03}^{+0.03} $&$ 7.55_{-0.81}^{+1.17}  $&$ 1.00_{-0.07}^{+0.11} $& 11.5 & 1.14 \\
Mar 20  &$ 0.85_{-0.31}^{+0.58} $&$ 1.48_{-0.39}^{+0.44} $&$ 13.07_{-0.70}^{+0.61} $&$ 4.58_{-0.08}^{+0.11} $&$ 10.6\pm0.1         $&$ 5.26_{-0.09}^{+0.10} $&$ 0.88\pm0.02          $&$ 7.78_{-0.59}^{+0.80}  $&$ 1.01_{-0.05}^{+0.07} $& 11.2 & 1.26 \\
Mar 21a &$ 0.57_{-0.19}^{+0.39} $&$ 1.19_{-0.36}^{+0.45} $&$ 12.82_{-0.63}^{+0.63} $&$ 4.68_{-0.04}^{+0.05} $&$ 10.5_{-0.1}^{+0.1} $&$ 5.61_{-0.08}^{+0.08} $&$ 0.82_{-0.02}^{+0.02} $&$ 8.51_{-0.37}^{+0.50}  $&$ 0.95_{-0.03}^{+0.04} $& 11.9 & 2.44 \\
Mar 21b &$ 0.52_{-0.19}^{+0.34} $&$ 1.13_{-0.40}^{+0.43} $&$ 12.44_{-0.79}^{+0.67} $&$ 4.64_{-0.08}^{+0.10} $&$ 10.6_{-0.2}^{+0.1} $&$ 5.66\pm0.11          $&$ 0.82\pm0.02          $&$ 7.86_{-0.66}^{+0.80}  $&$ 0.93_{-0.05}^{+0.06} $& 11.6 & 1.24 \\
Mar 22  &$ 0.85_{-0.30}^{+0.58} $&$ 1.61_{-0.37}^{+0.44} $&$ 12.16_{-0.46}^{+0.44} $&$ 4.61_{-0.04}^{+0.05} $&$ 10.5_{-0.1}^{+0.1} $&$ 5.39_{-0.08}^{+0.08} $&$ 0.87_{-0.02}^{+0.02} $&$ 7.93_{-0.36}^{+0.48}  $&$ 0.99_{-0.03}^{+0.04} $& 10.4 & 2.42 \\
Mar 27  &$ 0.85_{-0.38}^{+0.81} $&$ 1.90_{-0.47}^{+0.55} $&$ 8.85_{-0.34}^{+0.31}  $&$ 4.72_{-0.05}^{+0.06} $&$ 10.4_{-0.1}^{+0.1} $&$ 5.75_{-0.11}^{+0.10} $&$ 0.85_{-0.02}^{+0.02} $&$ 7.78_{-0.37}^{+0.51}  $&$ 1.08_{-0.03}^{+0.04} $& 7.64 & 2.09 \\
Mar 29  &$ 0.88_{-0.41}^{+0.64} $&$ 2.02_{-0.51}^{+0.45} $&$ 8.12_{-0.32}^{+0.24} $&$ 4.75_{-0.06}^{+0.07}  $&$ 10.3_{-0.2}^{+0.1} $&$ 6.24_{-0.13}^{+0.15} $&$ 0.82_{-0.02}^{+0.02} $&$ 8.34_{-0.46}^{+0.55}  $&$ 1.07_{-0.05}^{+0.05} $& 6.80 & 2.73 \\
Mar 31  &$ 0.71_{-0.28}^{+0.56} $&$ 1.92_{-0.42}^{+0.48} $&$ 7.04_{-0.27}^{+0.25} $&$ 4.75_{-0.07}^{+0.10}  $&$ 10.7_{-0.3}^{+0.2} $&$ 7.67_{-0.25}^{+0.24} $&$ 0.86\pm0.03          $&$ 7.93_{-0.73}^{+0.97}  $&$ 0.87_{-0.07}^{+0.08} $& 5.90 & 1.53 \\
Apr  1  &$ 1.08_{-0.56}^{+1.36} $&$ 2.23_{-0.60}^{+0.66} $&$ 7.10_{-0.58}^{+0.53} $&$ 4.58_{-0.17}^{+0.23}  $&$ 11.5_{-0.4}^{+0.3} $&$ 8.70_{-0.57}^{+0.72} $&$ 1.05\pm0.06          $&$ 5.13_{-2.41}^{+3.19} $&$ 0.56_{-0.18}^{+0.23} $& 5.28 & 0.76 \\
Apr  2  &$ 1.01_{-0.40}^{+0.71} $&$ 2.24_{-0.40}^{+0.42} $&$ 6.24_{-0.26}^{+0.25} $&$ 4.77_{-0.10}^{+0.18}  $&$ 11.4_{-0.3}^{+0.2} $&$ 10.84_{-0.72}^{+0.90}$&$ 1.04_{-0.04}^{+0.03} $&$ 3.58_{-2.27}^{+2.54} $&$ 0.26_{-0.08}^{+0.14} $& 4.89 & 1.16 \\
Apr  3  &$ 0.50_{-0.20}^{+0.37} $&$ 1.68_{-0.43}^{+0.46} $&$ 4.94_{-0.37}^{+0.39} $&$ 5.36_{-0.45}^{+0.56} $&$ 12.1_{-1.0}^{+0.8} $&$ 12.48_{-1.05}^{+1.08} $&$ 1.22\pm0.06          $&$ = W_1                $&$ \leq 0.69            $& 4.78 & 0.96 \\
Apr  4  &$ 0.70_{-0.26}^{+0.44} $&$ 2.03_{-0.37}^{+0.40} $&$ 4.84\pm0.19          $&$ 5.00_{-0.20}^{+0.22} $&$ 12.3_{-0.5}^{+0.4} $&$ 11.89_{-0.53}^{+0.54} $&$ 1.20\pm0.03          $&$ = W_1                $&$ \leq 0.54            $& 4.19 & 1.14 \\
Apr  5  &$ 0.37_{-0.14}^{+0.24} $&$ 1.60_{-0.38}^{+0.41} $&$ 4.02_{-0.26}^{+0.28} $&$ 5.14_{-0.34}^{+0.41} $&$ 12.7_{-0.7}^{+0.6} $&$ 12.05_{-0.87}^{+0.94} $&$ 1.27\pm0.06          $&$ = W_1                $&$ \leq 0.63            $& 3.77 & 0.67 \\
Apr  6  &$ 0.49_{-0.18}^{+0.31} $&$ 1.88_{-0.37}^{+0.40} $&$ 3.93\pm0.18          $&$ 5.19_{-0.25}^{+0.29} $&$ 11.9_{-0.6}^{+0.5} $&$ 12.36_{-0.59}^{+0.62} $&$ 1.21\pm0.03          $&$ = W_1                $&$ \leq 0.61            $& 3.47 & 1.21 \\
Apr  8  &$ 0.19_{-0.06}^{+0.09} $&$ 1.14_{-0.30}^{+0.32} $&$ 3.30_{-0.19}^{+0.20} $&$ 4.69_{-0.20}^{+0.22} $&$ 14.2\pm0.4         $&$ 10.89_{-0.60}^{+0.64} $&$ 1.32_{-0.06}^{+0.05} $&$ = W_1                $&$ \leq 0.49            $& 3.00 & 1.17 \\
Apr 10  &$ 0.13_{-0.04}^{+0.05} $&$ 0.82_{-0.26}^{+0.28} $&$ 2.79_{-0.24}^{+0.25} $&$ 4.25_{-0.22}^{+0.26} $&$ 14.9\pm0.4         $&$ 9.67_{-0.82}^{+0.87}  $&$ 1.17_{-0.10}^{+0.09} $&$ -                    $&$ -                    $& 2.39 & 0.87 \\
Apr 12  &$ 0.11_{-0.03}^{+0.04} $&$ 0.87_{-0.23}^{+0.24} $&$ 1.91_{-0.25}^{+0.29} $&$ 4.17_{-0.37}^{+0.48} $&$ 14.7_{-0.7}^{+0.5} $&$ 10.01_{-1.43}^{+1.52} $&$ 1.07_{-0.17}^{+0.16} $&$ -                    $&$ -                    $& 1.63 & 0.94 \\
Apr 14  &$ 0.13_{-0.03}^{+0.05} $&$ 1.22_{-0.21}^{+0.25} $&$ 1.32_{-0.16}^{+0.21} $&$ 4.61_{-0.67}^{+0.56} $&$ 14.9_{-0.6}^{+0.4} $&$ 10.76_{-1.29}^{+1.34} $&$ 1.32_{-0.18}^{+0.15} $&$ -                    $&$ -                    $& 1.27 & 1.00 \\
Apr 16  &$ 0.12_{-0.04}^{+0.10} $&$ 1.45_{-0.28}^{+0.42} $&$ 0.84_{-0.20}^{+0.34} $&$ 3.92_{-0.61}^{+0.75} $&$ 15.4_{-0.2}^{+0.4} $&$ 7.41_{-2.53}^{+2.32}  $&$ 1.03_{-0.39}^{+0.35} $&$ -                    $&$ -                    $& 0.69 & 1.19 \\
Apr 18  &$ 0.04_{-0.01}^{+0.03} $&$ 1.18_{-0.24}^{+0.36} $&$ 0.24_{-0.10}^{+0.12} $&$ 4.32_{-0.73}^{+1.75} $&$ 16.4_{-0.6}^{+0.6} $&$ 5.03_{-2.25}^{+3.16}  $&$ 1.06_{-0.39}^{+0.65} $&$ -                    $&$ -                    $& 0.34 & 1.36 \\
Apr 20  &$ 0.02_{-0.01}^{+0.03} $&$ 1.25_{-0.36}^{+0.56} $&$ 0.12_{-0.07}^{+0.09} $&$ 4.08_{-0.77}^{+1.96} $&$ 15.9_{-1.3}^{+1.2} $&$ 3.84_{-2.74}^{+5.35} $&$ 0.80_{-0.39}^{+1.06} $&$  -                    $&$ -                    $& 0.17 & 0.90 \\
\enddata

\tablenotetext{a}{$\times 10^{-3}$}
\tablenotetext{b}{The value of $E_{a2}$ was fixed at twice $E_{a1}$.}
\tablenotetext{c}{$10^{37}$ erg s$^{-1}$ in 3$-$50 keV}

\end{deluxetable}

\clearpage


\clearpage

\clearpage

\clearpage

\clearpage

\clearpage

\clearpage

\clearpage

\clearpage

\clearpage

\clearpage

\clearpage

\begin{thebibliography}{}
\bibitem[Alexander and Meszaros(1991)]{alm91} Alexander, S. G., \& Meszaros, P.  1991, \apj, 372, 565
\bibitem[Bildsten et al. (1997)]{bil97} Bildsten, L., Chakrabarty, D., Chiu, J., Finger, M. H., Koh, D. T., Nelson, R. W., Prince, T. A., Rubin, B. C., Scott, D. M., Stollberg, M., Vaughan, B. A., Wilson, C. A., \& Wilson, R. B., 1997, \apj, 113, 367
\bibitem[Burnard et al.(1991)]{bak91} Burnard, D. J., Arons, J., \& Klein, R. I. 1991, ApJ, 367, 575
\bibitem[Clark et al.(1994)]{cla94} Clark, G. W., Woo, J. W., \& Nagase, F. 1994, \apj, 422, 336
\bibitem[Coburn et al.(2001)]{Coburn2001} Coburn, W., Heindl, W. A., Gruber, D. E., Rothschild, R. E., Staubert, R., Wilms, J., \& Kreykenbohm, I. 2001, ApJ, 552, 738
\bibitem[Coburn et al.(2002)]{Coburn2002} Coburn, W., Heindl, W. A., Rothschild, R. E., Gruber, D. E., Kreykenbohm, I., Wilms, J., Kretschmar, P., \& Staubert, R. 2002, ApJ, 580, 394
\bibitem[Coburn et al.(2004)]{Coburn2004} Coburn, W., Kalemci, E., Kretschmar, P., Kreykenbohm, I., Rothschild, R., Staubert, R., \& Wilms, J. 2004, Atel, 381, 1
\bibitem[Finger et al.(1996)]{fin96} Finger, M. H., Wilson, R. B., \& Harmon, B. A. 1996, \apj, 459, 288
\bibitem[Heindl1999]{Heindl1999} Heindl, W. A., Coburn, W., Gruber, D. E., Pelling, M. R., Rothschild, R. E., Wilms, J., Pottschmidt, K., Staubert, R. 1999 , ApJ, 521, L49
\bibitem[Heindl2001]{Heindl2001} Heindl, W. A., Coburn, W., Gruber, D. E., Rothschild, R. E., Kreykenbohm, I., Wilms, J., \& Staubert, R. 2001 , ApJ, 563, L35
\bibitem[Jahoda1996]{Jahoda1996} Jahoda, K., Swank, J. H., Giles, A. B., Stark, M. J., Strohmayer, T., Zhang, W., \& Morgan, E. H. 1996, Proc. SPIE, 2808, 59
\bibitem[Kreykenbohm2002]{Kreykenbohm2002} Kreykenbohm, I., Coburn, W., Wilms, J., Kretschmar, P., Staubert, R., Heindl, W. A., \& Rothschild, E. 2002, A\&A, 395, 129
\bibitem[Kreykenbohm2005]{Kreykenbohm2005} Kreykenbohm, I., Mowlavi, N., Produit, N., Soldi, S., Walter, R., Dubath, P., Lubinski, P., Tr\"{u}mper, M., Coburn, W., Santangelo, A., Rothschild, R. E., \& Staubert, R. 2005, A\&A, 433, L45
\bibitem[Makishima1990]{Makishima1990} Makishima, K., Mihara, T., Ishida, M., Ohashi, T., Sakao, T., Tashiro, M., Tsuru, T., Kii, T., Makino, F., Murakami, T., Nagase, F., Tanaka, Y., Kunieda, H., Tawara, Y., Kitamoto, S., Miyamoto, S., Yoshida, A., \& Turner, M. J. L. 1990, ApJ, 365, L59 
\bibitem[Makishima1999]{Makishima1999} Makishima, K., Mihara, T., Nagase, F. \& Tanaka, Y. 1999, ApJ, 525, 978
\bibitem[M\'{e}sz\'{a}ros 1992]{Meszaros1992} M\'{e}sz\'{a}ros, P. 1992, High-Energy Radiation from Magnetized Neutron Stars (Chicago: Univ. Chicago Press) 
\bibitem[Mihara1990]{Mihara1990} Mihara, T., Makishima, K., Ohashi, T., Sakao, T., \& Tashiro, M. 1990, {\it Nature}, 346, 250
\bibitem[Mihara et al.\ 1991]{Mihara1991} Mihara, T., Ohashi, T., Makishima, K., Nagase, F., Kitamoto, S., \& Koyama, K. 1991, PASJ, 43, 501
\bibitem[Mihara1995]{Mihara1995} Mihara, T., Ph.D. thesis in University of Tokyo 1995
\bibitem[Mihara1998]{Mihara1998} Mihara, T., Makishima, K., \& Nagase, F. 1998, Adv. Space Res. 22, 987
\bibitem[Mihara2004]{Mihara2004} Mihara, T., Makishima, K., \& Nagase, F. 2004, ApJ, 610, 390
\bibitem[Nagase1989]{Nagase1989} Nagase, F., 1989, PASJ, 41, 1
\bibitem[Nagase1991]{Nagase1991} Nagase, F., Dotani, T., Tanaka, Y., Makishima, K., Mihara, T., Sakao, T., Tsunemi, H., Kitamoto, S., Tamura, K., Yoshida, A., \& Nakamura, H. 1991, ApJ, 375, L49
\bibitem[Negueruela2001]{Negueruela2001} Negueruela, I. \& Okazaki, A. T. 2001, A\&A, 369. 108
\bibitem[Okada2004]{Okada2004} Okada, Y., Niko, H., Makishima, K., Nakajima, M., Mihara, T., Terada, Y., Nagase, F., \& Tanaka, Y. 2004, Proc. 5th INTEGRAL Workshp
\bibitem[Orlandini1998]{Orlandini1998} Orlandini, M., dal Fiume, D., Frontera, F., Cusumano, G., del Sordo, S., Giarrusso, S., Piraino, S., Segreto, A., Guainazzi, M., \& Piro, L. 1998, A\&A, 332. 121
\bibitem[Remillard2004]{Remillard2004} Remillard, R., 2005, Atel, 371, 1
\bibitem[Romanova et al.\ 2004]{Romanova2004} Romanova, M. M., Ustyugova, G. V., Koldoba, A. V., \& Lovelace, R. V. E., 2004, \apj, 616, 151
\bibitem[Rose1979]{Rose1979} Rose, L. A., Marshall, F. E., Holt, S. S., Boldt, E. A., Rothschild, R. E., Serlemitsos, P. J., Pravdo, S. H., \& Kaluzienski, L. J. 1979, ApJ, 231, 919
\bibitem[Rothschild1998]{Rothschild1998} Rothschild, R. E., Blanco, P. R., Gruber, D. E., Heindl, W. A., MacDonald, D. R., Marsden, D. C., Pelling, M. R., Wayne, L. R., Hink, P. L. 1998, ApJ, 496, 538
\bibitem[Rybicki1979]{Rybicki1979} Rybicki, G. B., \& Lightman, A. P. 1979, Radiation Processed in Astrophysics, John Wiley \& Sons, Inc
\bibitem[Snatangelo1999]{Snatangelo1999} Santangelo, A., Segreto, A., Giarrusso, S., dal Fiume, D., Orlandini, M., Parmar, A. N., Oosterbroek, T., Bulik, T., Mihara, T., Campana, S., Israel, G. L., \& Stella, L. 1999, ApJ, 523, L85
\bibitem[Swank2004]{Swank2004} Swank, J., Remillard, R., \& Smith, E. Atel, 2004, 349, 1
\bibitem[Trumper1978]{Trumper1978} Tr\"{u}mper, J., Pietsch, W., Reppin, C., Voges, W., Staubert, R., \& Kendziorra, E. 1978, ApJ, 219, L105
\bibitem[Unger1998]{Unger1998} Unger, S. J., Roche, P., Negueruela, I., Ringwald, F. A., Lloyd, C., \& Coe, M. J. 1998, A\&A 336, 960
\bibitem[Wheaton1979]{Wheaton1979} Wheaton, W. A., Doty, J. P., Primini, F. A., Cooke, B. A., Dobson, C. A., Goldman, A., Hecht, M., Howe, S. K., Hoffman, J. A., \& Scheepmaker, A. 1979, {\it Nature}, 282, 240
\bibitem[White1983]{White1983} White, N., Swank, J., \& Holt, S.S.  1983, ApJ, 270, 711
\bibitem[Wilms1999]{Wilms1999} Wilms, J., Nowak, M. A., Dove, J. B., Fender, R. P., \& di Matteo, T. 1999, ApJ, 522, 460
\end{thebibliography}
\end{document}